\documentclass[tikz]{article}

\usepackage{PRIMEarxiv}

\usepackage[utf8]{inputenc} 
\usepackage[T1]{fontenc}    
\usepackage{hyperref}       
\usepackage{url}            
\usepackage{booktabs}       
\usepackage{amsfonts}       
\usepackage{nicefrac}       
\usepackage{microtype}      
\usepackage{lipsum}
\usepackage{fancyhdr}       
\usepackage{graphicx}       

\usepackage[title]{appendix}%
\usepackage{amsmath,amssymb,amsfonts}%
\usepackage{subfigure}
\usepackage{enumitem}
\usepackage{adjustbox}
\usepackage{amsmath}
\usepackage{array}
\usepackage{booktabs}
\usepackage{longtable}
\usepackage{caption}
\usepackage{multirow}
\usepackage{multicol}
\usepackage{rotating}
\usepackage{tabularx}
\usepackage{url}
\usepackage[table,xcdraw]{xcolor}
\usepackage{hyperref} 
\usepackage{balance}

\usepackage{mathrsfs}%
\usepackage{manyfoot}%
\usepackage{booktabs}%
\usepackage{algorithm}%
\usepackage{algorithmicx}%
\usepackage{algpseudocode}%
\usepackage{mathrsfs}
\usepackage{listings}%
\usepackage{float}
\usepackage{dsfont}
\usepackage{microtype}

\usepackage{pdflscape}
\usepackage{geometry}
\usepackage{fancyhdr} 
\fancypagestyle{mylandscape}{
\fancyhf{} 
\fancyfoot{
\makebox[\textwidth][r]{
  \rlap{\hspace{-.75cm}
    \smash{
      \raisebox{5.07in}{
        \rotatebox{90}{\thepage}}}}}}
}

\usepackage{tikz}

\usetikzlibrary{graphs, positioning, shapes.geometric, arrows.meta, fit, backgrounds}
\newcommand{\methodname}{HP-MOCD} 
\title{
    A High-Performance Evolutionary Multiobjective Community Detection Algorithm
}

\author{
Guilherme O. Santos$^1$, Lucas S. Vieira$^1$, Giulio Rossetti$^2$, \\ \textbf{Carlos H. G. Ferreira$^1$ and Gladston Moreira$^3$} \\
 \\
 $^1$Departamento de Computa\c{c}\~{a}o e Sistemas, Universidade Federal de Ouro
Preto,\\ Jo\~{a}o Monlevade, 35931-008, Minas Gerais, Brazil.\\
$^2$Istituto di Scienza e Tecnologie dell\'Informazione, National Research,\\
Council of Italy, Pisa, 56127, PI, Italy.\\
$^3$Computing Department, Universidade Federal de Ouro Preto,\\ Ouro
Preto, 35402-136, Minas Gerais, Brazil.\\
    \texttt{\{guilherme.os1, lucas.cv\}@aluno.ufop.edu.br}\\ \texttt{giulio.rossetti@isti.cnr.it, chgferreira@ufop.edu.br, gladston@ufop.edu.br} \\
}

\begin{document}
\maketitle

\begin{abstract}
Community detection in complex networks is fundamental across social, biological, and technological domains. While traditional single-objective methods like Louvain and Leiden are computationally efficient, they suffer from resolution bias and structural degeneracy. Multi-objective evolutionary algorithms (MOEAs) address these limitations by simultaneously optimizing conflicting structural criteria, however, their high computational costs have historically limited their application to small networks. We present HP-MOCD, a High-Performance Evolutionary Multiobjective Community Detection Algorithm built on Non-dominated Sorting Genetic Algorithm II (NSGA-II), which overcomes these barriers through topology-aware genetic operators, full parallelization, and bit-level optimizations—achieving theoretical $O(G \cdot N_p |V|)$ complexity. We conduct experiments on both synthetic and real-world networks. Results demonstrate strong scalability, with HP-MOCD processing networks of over 1,000,000 nodes while maintaining high quality across varying noise levels. It outperforms other MOEAs by more than 531 times in runtime on synthetic datasets, achieving runtimes as low as 57 seconds for graphs with 40,000 nodes on moderately powered hardware. Across 14 real-world networks, HP-MOCD was the only MOEA capable of processing the six largest datasets within a reasonable time, with results competitive with single-objective approaches. Unlike single-solution methods, HP-MOCD produces a Pareto Front, enabling individual-specific trade-offs and providing decision-makers with a spectrum of high-quality community structures. It introduces the first open-source \texttt{Python} MOEA library compatible with \texttt{networkx} and \texttt{igraph} for large-scale community detection.
\end{abstract}


\section{Introduction}\label{sec:introduction}

\begin{figure}[h!]
\centering
\begin{tikzpicture}[
    node_style/.style={circle, draw, fill, minimum size=0.5cm},
    community1_node/.style={node_style, fill=blue!40},
    community2_node/.style={node_style, fill=green!40},
    edge_style/.style={thick, gray},
    bridge_edge_style/.style={thick, red, dashed}
]

\foreach \angle in {0, 72, 144, 216, 288} {
    \node[community1_node] (c1-\angle) at (\angle:1.5cm) {};
}
\graph[edges={edge_style}] {
    (c1-0) -- (c1-72) -- (c1-144) -- (c1-216) -- (c1-288) -- (c1-0); 
    (c1-0) -- (c1-144);
    (c1-0) -- (c1-216);
    (c1-72) -- (c1-216);
    (c1-72) -- (c1-288);
    (c1-144) -- (c1-288);
};

\foreach \angle in {0, 72, 144, 216, 288} {
    \node[community2_node] (c2-\angle) at ([shift={(4.5cm,0)}]\angle:1.5cm) {};
}
\graph[edges={edge_style}] {
    (c2-0) -- (c2-72) -- (c2-144) -- (c2-216) -- (c2-288) -- (c2-0); 
    (c2-0) -- (c2-144);
    (c2-0) -- (c2-216);
    (c2-72) -- (c2-216);
    (c2-72) -- (c2-288);
    (c2-144) -- (c2-288);
};

\path[bridge_edge_style] (c1-0) edge[bend left=10] (c2-144);

\end{tikzpicture}
\caption{A conceptual illustration of community structure. The network is partitioned into two distinct communities (blue and green). The solid gray lines represent dense \textbf{intra-community edges}, which create cohesive groups. The single dashed red line represents a sparse \textbf{inter-community edge} (or bridge), highlighting the weak connectivity between the groups. The goal of community detection is to identify such partitions.}
\label{fig:conceptual_community_structure}
\end{figure}
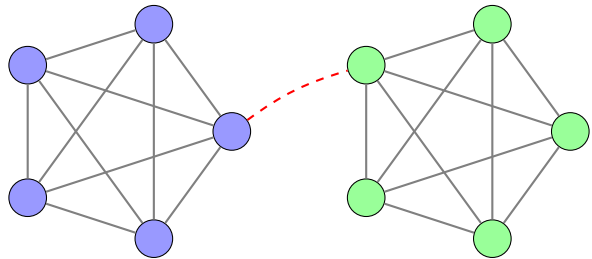

The study of complex systems, ubiquitous in social, technological, and biological domains, heavily relies on modeling their underlying connectivity patterns as complex networks \cite{Boccaletti2006}. Examples range from social media platforms and the World Wide Web to protein-protein interaction networks and critical infrastructures \cite{Ravasz2002}. A key characteristic of many such networks is their organization into communities, groups of nodes with dense internal connections and sparser connections to the rest of the network \cite{Newman2006}. \autoref{fig:conceptual_community_structure} provides a canonical example of community structure.
Identifying these communities is crucial for understanding network organization, function, and dynamics, making community detection a central research area in network science \cite{FORTUNATO:16, Flake2002, blondel2008, shi2010}. The task is commonly formulated as a combinatorial optimization problem, where the goal is to partition the network into clusters that maximize internal cohesion while minimizing interconnections between groups \cite{Newman2004}. Since this problem is NP-hard, most practical solutions rely on heuristic methods that optimize a predefined structural quality metric \cite{FORTUNATO201075}.

Among these, single-objective approaches are particularly widespread due to their efficiency and simplicity. However, they often oversimplify the multifaceted nature of real-world networks by focusing on a single structural criterion. Two of the most widely used single-objective methods are the Louvain algorithm \cite{blondel2008}, which greedily maximizes modularity, and its improved variant, Leiden \cite{traag2019}, which addresses issues such as disconnected communities through an enhanced refinement process. 

While effective and computationally efficient, these traditional heuristic methods exhibit notable limitations. Their reliance on a single metric like modularity introduces structural biases, including the well-documented resolution limit, which causes small but meaningful communities to be absorbed into larger ones \cite{Fortunato2007, Newman2004}. Modularity-based methods may also suffer from degeneracy, producing multiple partitions with similar scores but different topologies \cite{traag2019}. Although Leiden mitigates some of these issues by ensuring better-connected communities and more stable partitions, it still operates under a fixed optimization criterion. This inherently restricts their capacity to navigate the complex trade-offs that arise in real-world networks, where no single metric fully captures the richness of the underlying structure \cite{Azevedo2024}.

Recognizing the limitations of traditional heuristic methods, the field has progressively shifted towards multi-objective evolutionary algorithms (MOEAs) as a more expressive and flexible framework for community detection \cite{shaik2021evolutionary, dabaghi2025community}. Unlike single-metric approaches, MOEAs aim to optimize multiple, often conflicting, objectives simultaneously. This multi-criteria formulation enables a more nuanced characterization of community structures by capturing diverse structural features of the network, such as internal density, inter-community sparsity, and modularity. By design, these algorithms produce a Pareto front, i.e., a set of non-dominated solutions, offering practitioners a range of alternatives rather than a single fixed partition \cite{moreira:19}. This enables a more comprehensive exploration of the solution space, facilitating informed decisions based on application-specific trade-offs.

Early works such as MOGA-Net \cite{Pizzuti2009}, which jointly optimized a community score and a fitness-based criterion, and Shi-MOCD \cite{shi2010}, which explicitly targeted intra-community density and inter-community sparsity, demonstrated the potential of MOEAs to overcome structural limitations of modularity-based methods, including the resolution limit \cite{Fortunato2007}. Following these foundational works, the field has seen a range of proposals aimed at improving search efficiency and solution quality. In \cite{Liu2024}, the authors developed learning-based strategies to select nodes in each round of the multi-objective evolutionary process, called Local Search for Community Detection (LSCD). The work \cite{fu2024} introduces a continuous-encoding-based approach with a clustering coefficient in a multi-objective framework, called CECC-Net, which represents the community structure. Focusing on the challenge of detecting overlapping communities, the work \cite{yusupov2021} presents an MOEA that utilizes an ensemble of different initializations. Despite their theoretical appeal, the practical adoption of those MOEAs for large-scale community detection remains limited, with applications to synthetic and real-world networks of no more than 10,000 nodes. A key challenge is their computational cost. Maintaining and evolving a population of candidate solutions requires repeated dominance comparisons, Pareto sorting, and fitness evaluations, all of which can be expensive in networks with thousands or millions of nodes and edges. Many existing implementations exhibit time complexities on the order of $O(n^2)$ or worse, primarily due to operations involving pairwise comparisons and graph traversals \cite{shi2010, Pizzuti2012}. This makes them impractical for real-time or large-scale applications without specialized acceleration strategies.

More recent algorithms include KRM and CCM \cite{shaik2021evolutionary}, both based on the NSGA-III framework. KRM optimizes three objectives: Kernel-k-means, Ratio-cut, and Modularity using a reference-point guided selection mechanism, while CCM targets Community fitness, Community score, and Modularity. Another notable contribution is CDRME \cite{dabaghi2025community}, a four-phase evolutionary method that incorporates similarity-based initialization and random-walk-based seeding to balance cut density, ratio cut, and modularity across network scales.

Despite their theoretical appeal, the practical adoption of these MOEAs for large-scale community detection remains limited, with applications to synthetic and real-world networks of no more than 10,000 nodes. A key challenge is their computational cost. Maintaining and evolving a population of candidate solutions requires repeated dominance comparisons, Pareto sorting, and fitness evaluations, all of which can be expensive in networks with thousands or millions of nodes and edges. Many existing implementations exhibit time complexities on the order of $O(n^2)$ or worse, primarily due to operations involving pairwise comparisons and graph traversals \cite{shi2010, Pizzuti2012}. This makes them impractical for real-time or large-scale applications without specialized acceleration strategies.

Additionally, few implementations of MOEAs are publicly available for community detection, which limits accessibility for researchers and practitioners. Most available tools are either experimental, tied to specific datasets, or lack parallelization and optimization for modern hardware. Finally, designing effective genetic operators that preserve meaningful community structures while ensuring exploration of the vast combinatorial search space remains an open problem. Balancing exploitation and exploration in this context is especially difficult due to the discrete and topologically constrained nature of the community detection problem.

To address these challenges, this paper introduces the High-Performance Multi-Objective Community Detection (\methodname) algorithm. \methodname~is a scalable evolutionary algorithm based on the Non-Dominated Sorting Genetic Algorithm II (NSGA-II) framework \cite{nsga2Deb2002}, specifically engineered for accuracy and efficiency in large complex networks. It employs a parallel computing architecture and custom genetic operators that incorporate topological information to explore the solution space effectively. Instead of yielding a single partition, \methodname~produces a diverse set of Pareto-optimal solutions, each representing a different trade-off between conflicting structural objectives, thereby offering a more holistic understanding of network community structures. Moreover, this solution set can be flexibly analyzed under multiple evaluation metrics or selected according to the specific needs of a given application. Our primary contributions are as follows:

\begin{enumerate}[label=(\roman*)]
  \item We introduce a novel topology-aware crossover and mutation schemes that both refine community assignments and preserve population diversity, resulting in more accurate and structurally consistent community partitions.
  \item We prove that, for the sparse graphs typically encountered in practice (\(|E|=O(|V|)\)) with fixed population size \(N_p\), the per-generation cost reduces to $O\bigl(N_p\,|V|\bigr)$. Our experiments confirm runtimes remain well below worst-case bounds.
  \item We design a fully multi‐threaded NSGA-II architecture that exploits modern multi‐core processors, drastically reducing runtime on large‐scale networks, .
  \item We provide a Free/Libre and Open-Source Software (FLOSS) implementation of \methodname, enabling researchers and practitioners to apply, extend, and benchmark our approach.
\end{enumerate}

The remainder of this paper is structured as follows. Section~\ref{sec:background} reviews foundational concepts in community detection, including definitions and key modeling assumptions. Section~\ref{sec:relatedWork} surveys prior work, covering both traditional algorithms and recent multi-objective approaches. Section~\ref{sec:proposed_algorithm} formalizes the multi-objective problem and presents the HP-MOCD algorithm in detail, including its architecture, genetic operators, and parallel execution model. Section~\ref{sec:experiments} describes the experimental setup and reports a comparative evaluation against several baselines on synthetic and real-world networks. Finally, Section~\ref{sec:conclusion} discusses the implications, limitations, and potential directions for future research.

\section{Background}
\label{sec:background}

Community detection is a fundamental problem in network science, central to understanding the structure and function of complex systems. This section first introduces the foundational definitions and modeling assumptions adopted in this work. We then briefly review traditional heuristic methods, with emphasis on modularity-based algorithms such as Louvain and Leiden, analyzing their performance and limitations. Finally, we present an overview of multi-objective community detection, discussing its theoretical foundations, algorithmic strategies, and the key contributions from the literature that inform the design of our proposed method.

\subsection{The Definition of Community}

Community structure is a central concept in the analysis of complex networks, referring to the tendency of nodes to organize into groups with dense internal connections and relatively sparse connections to the rest of the network. While no universally accepted definition exists, one of the most common structural formulations is based on the relative density of intra- and inter-community links \cite{Radicchi2004}. In this perspective, communities are understood as subgraphs in which internal connectivity is significantly stronger than external connectivity.

Several formulations of community detection have emerged in the literature, depending on the nature of the data and the intended application \cite{FORTUNATO201075}. Among the most common is node clustering, where the goal is to assign each node to a community. Other approaches include edge clustering, which focuses on grouping edges rather than nodes, and overlapping or fuzzy clustering, in which nodes may belong to multiple communities, either to the same degree or with varying levels of association. More specialized formulations incorporate additional structural or semantic information \cite{fortunato:09}. Attributed clustering, for example, leverages both network topology and node-level features, while bipartite clustering targets networks composed of two distinct node types. Temporal clustering extends the task to dynamic networks, capturing how communities evolve over time. In directed acyclic graphs (DAGs), antichain clustering is used to identify groups based on partial order relationships \cite{Rossetti2019CDLib}.

In this work, we focus on node clustering under a crisp partitioning model, also referred to as hard clustering, where each node is assigned to exactly one community. This assumption of mutually exclusive group membership simplifies the structural analysis and remains the most widely adopted formulation in the evaluation of community detection algorithms. We further restrict our scope to undirected and unweighted networks, which are commonly used in benchmarks and allow clear structural interpretations based solely on topology.

Formally, let \(G = (V, E)\) be an undirected and unweighted graph, where \(V\) is a finite, non-empty set of nodes and \(E \subseteq \{\{u,v\} \mid u,v \in V, u \ne v\}\) is the set of undirected edges.

\begin{equation}
    \bigcup_{i=1}^{k} C_i = V \quad \text{and} \quad C_i \cap C_j = \emptyset \quad \forall \ i \neq j
\end{equation}

\subsection{Multi-Objective Optimization}
\label{MO-Algorithms}

A multi-objective optimization problem (MOP) involves simultaneously optimizing two or more, often conflicting, objective functions. Improving one objective may lead to the deterioration of another, making it impractical to obtain a single solution that is optimal for all objectives. Instead, MOP aims to find a set of solutions that represent optimal trade-offs among the objectives. In this study, optimization is approached as the minimization of objective values.

Formally, a MOP can be stated as \cite{Ehr:00}:
\begin{align}
\mbox{Minimize} ~~\mathbf{f}(\mathbf{x}) = (f_1(\mathbf{x}), f_2(\mathbf{x}), \ldots, f_m(\mathbf{x})) 
\mbox{ subject to }
\mathbf{x} \in \Omega
\end{align}
Here, $\mathbf{x} = (x_1, x_2, \ldots, x_n)$ is a vector of $n$ decision variables residing in the $n$-dimensional decision space $\mathbb{R}^n$. The set $\Omega \subseteq \mathbb{R}^n$ represents the feasible region of the decision space, containing all permissible solutions. In the context of community detection, $\mathbf{x}$ typically represents a candidate partitioning of the network's nodes into communities. The function $\mathbf{f}: \Omega \to \mathbb{R}^m$ maps the decision space to the $m$-dimensional objective space, where each $f_i(\mathbf{x})$ is an objective function to be minimized. The image of the feasible set, $\mathbf{f}(\Omega) = \{ \mathbf{f}(\mathbf{x}) \mid \mathbf{x} \in \Omega \}$, is the set of all possible objective vectors in the objective space $\mathbb{R}^m$. The problem can also be subject to additional constraints in $\Omega$ and/or $\mathbf{f}(\Omega)$.

Since the objective space $\mathbb{R}^m$ (for $m > 1$) is only partially ordered, the concept of dominance is used to compare solutions. Let $\mathbf{a} = (a_1, a_2, \ldots, a_m)$ and $\mathbf{b} = (b_1, b_2, \ldots, b_m)$ be two objective vectors in $\mathbb{R}^m$. The vector $\mathbf{a}$ is said to dominate $\mathbf{b}$ (denoted as $\mathbf{a} \prec \mathbf{b}$) if and only if $a_i \leq b_i$ for all $i \in \{1, \ldots, m\}$ and $a_j < b_j$ for at least one $j \in \{1, \ldots, m\}$.
A solution $\mathbf{x}^* \in \Omega$ is Pareto optimal if there is no other solution $\mathbf{x} \in \Omega$ such that $\mathbf{f}(\mathbf{x})$ dominates $\mathbf{f}(\mathbf{x}^*)$. In other words, $\mathbf{x}^*$ is Pareto optimal if its corresponding objective vector $\mathbf{f}(\mathbf{x}^*)$ is non-dominated.

The set of all non-dominated objective vectors is known as the Pareto Front ($PF$), and the set of all Pareto optimal decision vectors is the Pareto Optimal set ($PO$):
\begin{align}
	PF & = \{ \mathbf{y} \in \mathbf{f}(\Omega) \mid \nexists \mathbf{y}' \in \mathbf{f}(\Omega) \text{ such that } \mathbf{y}' \prec \mathbf{y} \} \\
	PO & = \{ \mathbf{x} \in \Omega \mid \mathbf{f}(\mathbf{x}) \in PF \}
\end{align}
The $PF$ represents the set of optimal trade-off solutions where no objective can be improved without degrading at least one other objective \cite{moreira:19}.

\section{Related Work}\label{sec:relatedWork}

Community detection is a vast field. Here, we compare and contrast \methodname~with the two main classes of algorithms relevant to our work: traditional algorithms and other MOEAs.

\subsection{Traditional Algorithms}
\label{sec:SO-Algorithms}

Traditional approaches to community detection in complex networks often rely on the optimization of a structural quality function, the most prominent being modularity. These methods aim to partition a network such that the density of edges within communities is significantly higher than between them. While modularity-based algorithms remain dominant due to their efficiency and simplicity, several alternative heuristics have also shaped the field. Among the earliest and most influential is the Girvan–Newman algorithm, which detects communities by iteratively removing edges with the highest betweenness centrality, gradually decomposing the network into disconnected components \cite{girvan2002}. Despite its conceptual elegance, this method suffers from high computational cost and limited scalability.

Another prominent strategy is label propagation, a fast heuristic where each node adopts the most frequent label among its neighbors. This process converges to a partition where nodes in the same community share the same label \cite{raghavan2007}. Label propagation is highly scalable but can yield unstable results and lacks a clear objective function, making its outcomes harder to interpret. Among modularity-based methods, the Louvain algorithm proposed by \cite{blondel2008} has become one of the most widely used due to its balance between accuracy and computational performance. It seeks to optimize the modularity metric~\cite{newman2006modularity}, which measures the quality of a partition by comparing the density of edges within communities to the expected density in a randomized null model that preserves the node degrees. In essence, modularity quantifies how much more densely connected the nodes are within communities than would be expected by chance. The modularity score theoretically ranges from \(-\frac{1}{2}\) to 1, where higher values indicate a stronger community structure, and values near zero or negative suggest that the detected communities are no better—or even worse—than random.

%


Despite its popularity and efficiency, the Louvain algorithm exhibits several well-known limitations that affect both the quality and stability of the resulting partitions \cite{Fortunato2007}. Some works have attempted to address these limitations \cite{Wang2019}. One key issue is the production of disconnected communities, a consequence of how nodes are greedily reassigned based solely on local modularity gain without enforcing internal connectivity constraints \cite{FORTUNATO:16}. This can lead to communities that are not internally cohesive, violating a fundamental structural assumption in community detection. Furthermore, Louvain is sensitive to the order in which nodes are processed during the local moving phase. Because the algorithm applies greedy updates, different node orderings may lead to different partitions with comparable modularity scores, introducing a level of non-determinism that compromises reproducibility.

A more fundamental limitation lies in the modularity function itself, particularly its susceptibility to the resolution limit problem. As demonstrated by \cite{Fortunato2007}, modularity optimization tends to favor large communities and may fail to identify smaller but structurally meaningful groups, especially in networks with heterogeneous community sizes. This limitation arises because modularity evaluates gains relative to a global null model, which biases the optimization toward partitions that absorb smaller communities into larger ones. Additionally, the modularity landscape often contains many local maxima, leading to a high degree of degeneracy in the solution space. This means that multiple partitions with very different community structures can yield similar modularity values, complicating the interpretation and comparison of results across runs.

To address some of these issues, the Leiden algorithm was introduced by \cite{traag2019}. It preserves the modularity optimization framework but improves upon Louvain by enforcing connectedness within communities and refining the aggregation process. The algorithm proceeds in three stages. The first is local moving, in which nodes are reassigned to neighboring communities to improve a quality function. The second is refinement, which ensures that all communities remain internally connected, splitting any that are not. The final stage is aggregation, where communities are collapsed into single nodes, and the process restarts on the simplified graph.

While the Leiden algorithm was originally introduced with support for multiple objective functions, it often employs the Constant Potts Model (CPM) instead of modularity. CPM defines community quality in terms of internal edge density, controlled by a resolution parameter \( \gamma \), and is given by:
\begin{equation}
H = \sum_c \left( e_c - \gamma \frac{n_c(n_c - 1)}{2} \right)
\end{equation}

In this expression, \( e_c \) denotes the number of internal edges in community \( c \), \( n_c \) is the number of nodes in that community, and \( \gamma \) sets the minimum internal density threshold. By tuning \( \gamma \), the algorithm can control the granularity of the resulting partition, identifying larger or smaller communities accordingly. Unlike modularity, CPM does not suffer from the same theoretical resolution limit because it is a local objective function, and its optimization does not depend on global graph properties such as the total number of edges.

Nevertheless, in practical applications, CPM introduces its own set of challenges. The performance of the algorithm can vary significantly with the choice of \( \gamma \), which often requires careful, problem-specific calibration. Furthermore, when Leiden is configured to use modularity as its objective function, as is commonly done for compatibility with existing benchmarks, it inherits many of the same limitations as Louvain, including the resolution limit and modularity degeneracy. In either case, whether using modularity or CPM, Leiden remains fundamentally a single-objective optimization method.

Algorithms such as Louvain and Leiden represent the current state of practice in scalable, modularity-based community detection. Their combination of simplicity, computational efficiency, and availability in widely used libraries has made them widely adopted solutions in many real-world applications.  Despite this practical success, their reliance on a single optimization criterion inherently limits their expressiveness. In particular, they are not designed to capture multiple, potentially conflicting structural properties simultaneously, nor do they offer the user flexibility in selecting solutions that trade off different qualities such as compactness, separation, or size balance. These methods produce only a single partition per execution, often overlooking alternative high-quality solutions that might better suit specific applications or structural goals. These limitations motivate the investigation of multi-objective optimization strategies, which aim to overcome the rigidity of single-objective heuristics by generating diverse sets of non-dominated solutions. In the following section, we review some of the recent advances in this area and highlight their potential to provide richer structural insights and greater flexibility for real-world community detection.

\subsection{Multi-Objective Evolutionary Algorithms}

The limitations of single-objective methods motivated the development of Multi-Objective Evolutionary Algorithms (MOEAs) for community detection. Unlike single-metric approaches, MOEAs optimize multiple, often conflicting, objectives simultaneously, such as internal density and external sparsity. This approach yields a Pareto front\footnote{The proximity of the set of solutions obtained by an MOEA to the true $PF$ is a measure of the algorithm's convergence.}, a set of non-dominated solutions—offering a range of robust alternatives rather than a single, potentially biased partition \cite{moreira:19, coello2007}. The use of MOEAs has become a prominent strategy due to their population-based nature, which is well-suited for exploring the complex and combinatorial solution space of community detection \cite{Deb2001, Handl2007}.

Among the earliest applications of evolutionary multi-objective optimization in this domain is \textbf{MOGA-Net}, introduced by  \cite{Pizzuti2009}. Her approach uses genetic algorithms to evolve network partitions that optimize both a community score and modularity. Although pioneering, the method remains dependent on modularity as a selection criterion, which reintroduces resolution bias. Additionally, the algorithm's computational cost is not thoroughly characterized, leaving its applicability to large networks uncertain.

The Multi‐Objective Community Detection (MOCD) algorithm by \cite{shi2010} jointly minimizes two objectives related to intra-community density and inter-community sparsity. MOCD employs the PESA-II evolutionary framework and locus-based operators. However, its selection strategy and operators lead to high computational overhead, with a complexity of $O(gs^2(m+n))$, making it impractical for large networks.

More recent research has continued to explore the potential of MOEAs, introducing more sophisticated objective functions and hybrid strategies. The work by \cite{shaik2021evolutionary}, introduced a customized NSGA-III framework to simultaneously optimize three objectives (e.g., Kernel k-means, Ratio-cut, and Modularity). Their contribution demonstrated the feasibility of using many-objective optimization to find community structures that represent a more nuanced trade-off between different structural properties. \cite{dabaghi2025community} proposed a hybrid method that first uses random walks to generate an initial set of high-quality partitions, which are then refined by a multi-objective evolutionary algorithm. This approach aims to accelerate convergence by seeding the evolutionary process in a more promising region of the search space.

\section{The HP-MOCD Algorithm}
\label{sec:proposed_algorithm}


We now describe the High-Performance Multi-Objective Community Detection (\methodname) algorithm. Our method integrates the NSGA-II optimization framework with a parallel architecture and custom genetic operators designed for network topology.
The implementation is written in \href{https://www.rust-lang.org/}{Rust} for performance and exposed to Python via \href{https://pyo3.rs/v0.24.0/}{PyO3}. The full source code is publicly available on \href{https://oliveira-sh.github.io/pymocd/}{GitHub} and the documention and usage are available on the \href{https://oliveira-sh.github.io/pymocd/}{Official Documentation}. Algorithm~\autoref{alg:hp-mocd-main} and \autoref{fig:hpmocd_flowchart} summarizes the \methodname's algorithm.

\subsection{Overview and Design Rationale}

Traditional single-objective approaches often suffer from resolution limit problems and can become trapped in local optima, as discussed in Section~\ref{sec:background}. To address these limitations, we need to use metrics that discover the entire Pareto front of solutions, ranging from coarse to fine partitions without biasing toward any particular resolution level.

Given a graph $G = (V, E)$ with node set $V$ and edge set $E$, let $C = \{c_1, c_2, \ldots, c_k\}$ represent a partition of $V$ into $k$ communities. The set of all valid partitions of $V$ is denoted as the feasible space $\Omega = \{C \mid C \text{ is a valid partition of } V\}$.

Following the formulation of \cite{shi2010}, we decompose the community detection problem into two complementary objectives that capture the essential trade-offs in community structure. Our method seeks to find partitions that simultaneously minimize both objective functions:
\begin{equation}
\label{eq:objectives}
f_{1}(C) = 1 - \sum_{c \in C}\frac{\lvert E(c)\rvert}{m}, 
\quad
f_{2}(C) = \sum_{c \in C}\Bigl(\frac{\sum_{v \in c}\deg(v)}{2m}\Bigr)^{2}
\end{equation}
where $m = |E|$ denotes the total number of edges, $\deg(v)$ represents the degree of node $v$, and $|E(c)|$ is the number of internal edges within community $c$.

The first objective function $f_1$ (or $intra(C)$) measures the fraction of edges that are \emph{not} internal to communities, effectively penalizing partitions with poor internal cohesion. Minimizing $f_1$ encourages the formation of densely connected communities. The second objective function $f_2$ (or $inter(C)$) represents the sum of squared relative community sizes (measured by total degree), which penalizes the formation of overly large communities. Minimizing $f_2$ promotes more balanced partitions and prevents the emergence of dominant communities that absorb an excessive number of nodes.

These objectives are inherently conflicting; optimizing for strong internal cohesion (low $f_1$) often leads to fewer, larger communities, while optimizing for balanced community sizes (low $f_2$) typically results in many smaller communities with potentially weaker internal structure.

The problem is then stated as:
\begin{align}\label{eq:problem}
\min_{C \,\in\, \Omega}\;\mathbf{f}(C)
\;=\;
\min_{C \,\in\, \Omega}
\;\bigl(f_{1}(C),\,f_{2}(C)\bigr).
\end{align}

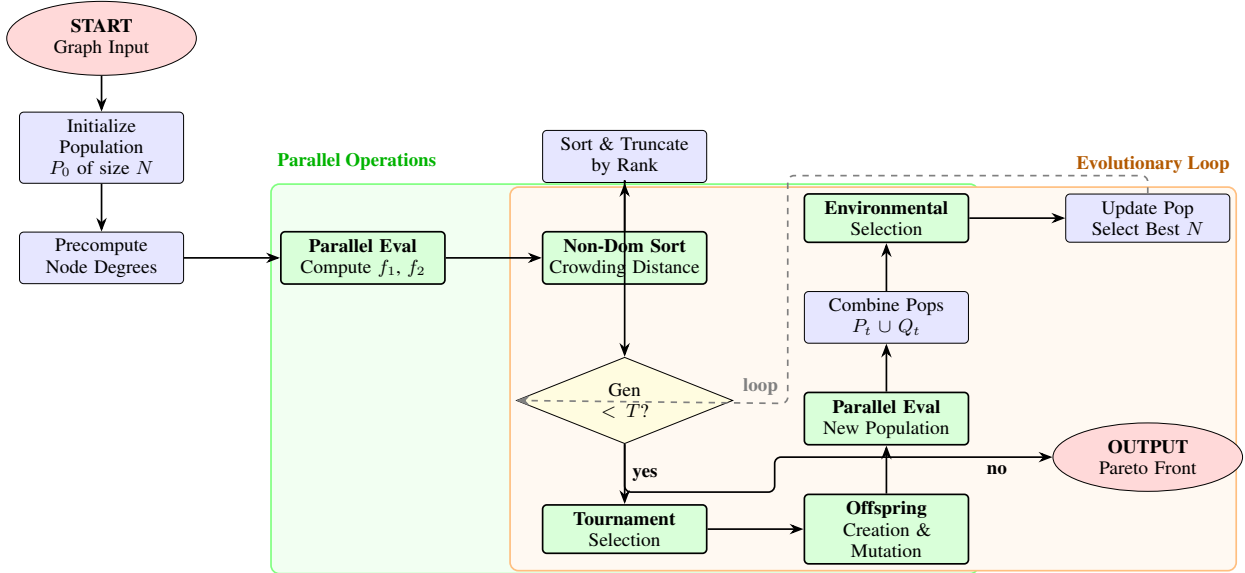
\begin{figure}[!ht]
\centering
\resizebox{\textwidth}{!}{
\begin{tikzpicture}[
    node distance=0.8cm and 1.0cm,
    every node/.style={font=\small},
    block/.style={rectangle, draw, rounded corners=2pt, text width=2.5cm, align=center, minimum height=0.7cm, fill=blue!10},
    parallel/.style={rectangle, draw, rounded corners=2pt, text width=2.5cm, align=center, minimum height=0.7cm, fill=green!15, thick},
    decision/.style={diamond, draw, aspect=2.5, text width=1.8cm, align=center, inner sep=0.05cm, fill=yellow!15},
    terminator/.style={ellipse, draw, text width=2.0cm, align=center, minimum height=0.6cm, fill=red!15},
    arrow/.style={thick,->,>=Stealth},
    loop/.style={thick,->,>=Stealth}
]

\node (start)     [terminator] {\textbf{START}\\Graph Input};
\node (init)      [block, below=0.6cm of start] {Initialize Population\\$P_0$ of size $N$};
\node (precomp)   [block, below=of init] {Precompute\\Node Degrees};

\node (eval1)     [parallel, right=1.6cm of precomp] {\textbf{Parallel Eval}\\Compute $f_1$, $f_2$};

\node (sort1)     [parallel, right=1.6cm of eval1] {\textbf{Non-Dom Sort}\\Crowding Distance};
\node (sort_pop)  [block, above=of sort1] {Sort \& Truncate\\by Rank};
\node (gen_check) [decision, below=1.2cm of sort1] {Gen\\$< T$?};
\node (selection) [parallel, below=1.0cm of gen_check] {\textbf{Tournament}\\Selection};

\node (offspring) [parallel, right=1.6cm of selection] {\textbf{Offspring}\\Creation \& Mutation};
\node (eval2)     [parallel, above=of offspring] {\textbf{Parallel Eval}\\New Population};
\node (combine)   [block, above=of eval2] {Combine Pops\\$P_t \cup Q_t$};
\node (env_sel)   [parallel, above=of combine] {\textbf{Environmental}\\Selection};

\node (update)    [block, right=1.6cm of env_sel] {Update Pop\\Select Best $N$};
\node (final)     [terminator, below=3.0cm of update] {\textbf{OUTPUT}\\Pareto Front};

\draw [arrow] (start) -- (init);
\draw [arrow] (init) -- (precomp);
\draw [arrow] (precomp) -- (eval1);
\draw [arrow] (eval1) -- (sort1);
\draw [arrow] (sort1) -- (sort_pop);

\draw [arrow] (sort_pop) -- ++(0,-0.4) -| (gen_check.north);

\draw [arrow] (gen_check.south) -- node[midway, right, font=\footnotesize] {\textbf{yes}} (selection.north);
\draw [arrow] (selection) -- (offspring);
\draw [arrow] (offspring) -- (eval2);
\draw [arrow] (eval2) -- (combine);
\draw [arrow] (combine) -- (env_sel);
\draw [arrow] (env_sel) -- (update);

\draw [arrow, dashed, gray, rounded corners=3pt] 
    (update.north) -- ++(0,0.3) -- ++(-6,0) -- ++(0,-3.8) -- 
    node[pos=0.1, above, font=\footnotesize] {\textbf{loop}} (gen_check.west);

\draw [arrow, rounded corners=3pt] 
    (gen_check.south) -- ++(0,-0.8) -- ++(2.5,0) |- 
    node[pos=0.9, below, font=\footnotesize] {\textbf{no}} (final.west);

\begin{scope}[on background layer]
    \node [draw=green!50, thick, rounded corners=4pt, fill=green!5, 
           fit={(eval1)(sort1)(selection)(offspring)(eval2)(env_sel)},
           inner sep=0.15cm] (parallel_group) {};
    \node [above=0.1cm of parallel_group.north west, anchor=south west, 
           font=\footnotesize, text=green!70!black] {\textbf{Parallel Operations}};
    
    \node [draw=orange!50, thick, rounded corners=4pt, fill=orange!5, 
           fit={(gen_check)(selection)(offspring)(eval2)(combine)(env_sel)(update)},
           inner sep=0.1cm] (loop_group) {};
    \node [above=0.1cm of loop_group.north east, anchor=south east, 
           font=\footnotesize, text=orange!70!black] {\textbf{Evolutionary Loop}};
\end{scope}

\end{tikzpicture}}
\caption{Flowchart of the HP-MOCD algorithm. Green-highlighted boxes indicate parallel operations that enhance computational efficiency. The algorithm iteratively evolves populations through selection, crossover, mutation, and environmental selection until convergence.}
\label{fig:hpmocd_flowchart}
\end{figure}

\autoref{fig:hpmocd_flowchart} illustrates the diagram of the proposed HP-MOCD algorithm, the initialization, evaluation, and evolutionary selection phases. Parallelism is applied during the evaluation of objectives and
sorting operations, enhancing scalability. The evolutionary loop iteratively generates offspring, combines populations, and applies environmental selection until the maximum number of generations is reached.
....

Our \methodname ~algorithm addresses this optimization problem, employing the NSGA-II framework and parallel architecture. The overall process is outlined in Algorithm \autoref{alg:hp-mocd-main}. It begins by creating an initial population of random partitions, each representing a possible way to partition the graph into communities. Then, through successive generations, it employs genetic operators (selection, crossover, and mutation) to evolve the population, preserving the best non-dominated solutions. 
\begin{algorithm}[!ht]
\small
\caption{High-Performance Multi-Objective Community Detection (HP-MOCD)}\label{alg:hp-mocd-main}
\begin{algorithmic}[1]
\Procedure{HP-MOCD}{Graph $G(V,E)$, Population Size $N_p$, Max Generations $T$, Crossover Probability $C_P$, Mutation Probability $M_P$, Ensemble Size $E_S$}
    \State \textbf{Input:} Undirected unweighted graph $G = (V, E)$ with $|V|$ nodes and $|E|$ edges
    \State \textbf{Output:} Pareto front $\mathcal{F}_1$ of non-dominated community partitions
    
    \State \Comment{\textbf{Phase 1: Initialization and Preprocessing}}
    \State $degrees \gets \text{PrecomputeDegrees}(G)$ \Comment{$O(|V|)$ cache for $\deg(v), \forall v \in V$}
    \State $m \gets |E|$ \Comment{Total number of edges}
    \State $P^{(0)} \gets \text{InitializePopulation}(G, N_p)$ \Comment{Random hash-map partitions}
    
    \State \Comment{\textbf{Phase 2: Initial Population Evaluation}}
    \State \textbf{parallel for each} $p \in P^{(0)}$ \textbf{do}
        \State \quad $\text{EvaluateIndividual}(p, G, degrees)$ \Comment{Compute $(f_1, f_2)$}
    \State \textbf{end parallel}
    
    \State \Comment{\textbf{Phase 3: Evolutionary Optimization Loop}}
    \For{$gen \gets 1$ \textbf{to} $T$}
        \State \Comment{\texttt{Multi-objective selection and ranking}}
        \State $\text{FastNonDominatedSort}(P^{(gen-1)})$ \Comment{$O(N_p \log N_p)$ for 2 objectives}
        \State $\text{CrowdingDistanceAssignment}(P^{(gen-1)})$ \Comment{Diversity preservation}
        \State $\text{SortByRankAndDistance}(P^{(gen-1)}, N_p)$ \Comment{Pareto-based ranking}
        
        \State \Comment{\texttt{Parallel offspring generation}}
        \State \textbf{parallel map} $i \gets 1$ \textbf{to} $N_p$ \textbf{do}
            \State \quad $\text{parents} \gets \text{TournamentSelection}(P^{(gen-1)}, E_S, k=2)$
            \State \quad $\text{child} \gets \text{TopologyAwareCrossover}(\text{parents}, C_P)$ \Comment{Algorithm~\ref{alg:crossover}}
            \State \quad $\text{TopologyAwareMutation}(\text{child}, G, M_P)$ \Comment{Section~\ref{sec:mutation}}
            \State \quad \textbf{return} $\text{child}$
        \State \textbf{end parallel} $\rightarrow Q^{(gen)}$
        
        \State \Comment{\texttt{Parallel offspring evaluation}}
        \State \textbf{parallel for each} $q \in Q^{(gen)}$ \textbf{do}
            \State \quad $\text{EvaluateIndividual}(q, G, degrees)$ 
        \State \textbf{end parallel}
        
        \State \Comment{\texttt{Environmental selection}}
        \State $R^{(gen)} \gets P^{(gen-1)} \cup Q^{(gen)}$ \Comment{Combine parent and offspring populations}
        \State $P^{(gen)} \gets \text{EnvironmentalSelection}(R^{(gen)}, N_p)$ \Comment{Select best $N_p$ individuals}
    \EndFor
    
    \State \Comment{\textbf{Phase 4: Pareto Front Extraction and Solution Selection}}
    \State $\mathcal{F}_1 \gets \{p \in P^{(T)} \mid \text{rank}(p) = 1\}$ \Comment{Extract Pareto front}
    \State $p^* \gets \text{SelectSolution}(\mathcal{F}_1)$ \Comment{Algorithm~\ref{alg:selection}}
    \State \Return $\mathcal{F}_1, p^*$ \Comment{Return Pareto front and recommended solution}
\EndProcedure
\end{algorithmic}
\end{algorithm}

\subsection{Solution Representation}
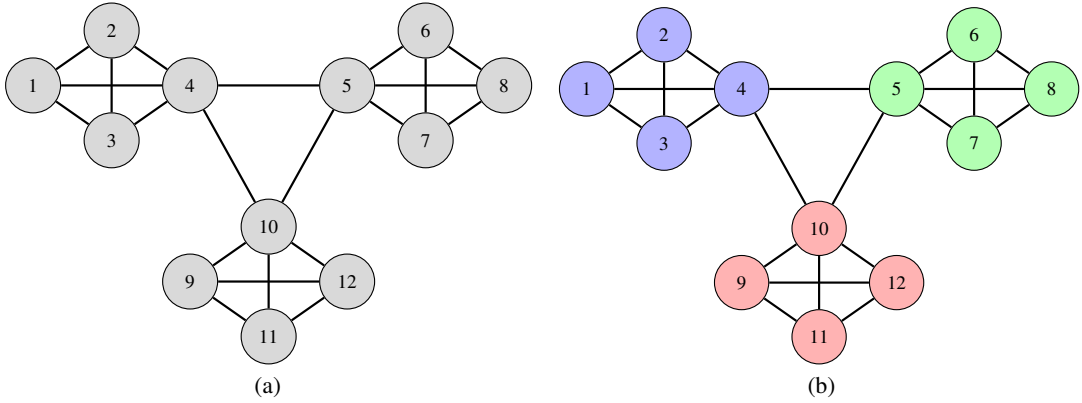
\begin{figure}[!ht] 
    \centering
        \subfigure[]{
        \resizebox{0.43\textwidth}{!}{
\begin{tikzpicture}[
    node_style/.style={circle, draw, minimum size=0.7cm, font=\scriptsize}, 
    community1_node/.style={node_style, fill=gray!30},
    community2_node/.style={node_style, fill=gray!30},
    community3_node/.style={node_style, fill=gray!30}, 
    edge_style/.style={thick}
]

\node[community1_node] (c1n1) at (0,2.5) {1};
\node[community1_node] (c1n2) at (1,3.2) {2};
\node[community1_node] (c1n3) at (1,1.8) {3};
\node[community1_node] (c1n4) at (2,2.5) {4};

\node[community2_node] (c2n1) at (4,2.5) {5};
\node[community2_node] (c2n2) at (5,3.2) {6};
\node[community2_node] (c2n3) at (5,1.8) {7};
\node[community2_node] (c2n4) at (6,2.5) {8};

\node[community3_node] (c3n1) at (2,0) {9};
\node[community3_node] (c3n2) at (3,0.7) {10};
\node[community3_node] (c3n3) at (3,-0.7) {11};
\node[community3_node] (c3n4) at (4,0) {12};

\graph[edges={edge_style}] {
    (c1n1) -- (c1n2);
    (c1n1) -- (c1n3);
    (c1n1) -- (c1n4);
    (c1n2) -- (c1n3); 
    (c1n2) -- (c1n4);
    (c1n3) -- (c1n4);
};

\graph[edges={edge_style}] {
    (c2n1) -- (c2n2);
    (c2n1) -- (c2n3);
    (c2n1) -- (c2n4);
    (c2n2) -- (c2n3); 
    (c2n2) -- (c2n4);
    (c2n3) -- (c2n4);
};

\graph[edges={edge_style}] {
    (c3n1) -- (c3n2);
    (c3n1) -- (c3n3);
    (c3n1) -- (c3n4);
    (c3n2) -- (c3n3); 
    (c3n2) -- (c3n4);
    (c3n3) -- (c3n4);
};

\path[edge_style] (c1n4) edge (c2n1); 
\path[edge_style] (c1n4) edge (c3n2); 
\path[edge_style] (c2n1) edge (c3n2); 


\end{tikzpicture}
}
} 
        \subfigure[]{
        \resizebox{0.43\textwidth}{!}{
\begin{tikzpicture}[
    node_style/.style={circle, draw, minimum size=0.7cm, font=\scriptsize}, 
    community1_node/.style={node_style, fill=blue!30},
    community2_node/.style={node_style, fill=green!30},
    community3_node/.style={node_style, fill=red!30}, 
    edge_style/.style={thick}
]

\node[community1_node] (c1n1) at (0,2.5) {1};
\node[community1_node] (c1n2) at (1,3.2) {2};
\node[community1_node] (c1n3) at (1,1.8) {3};
\node[community1_node] (c1n4) at (2,2.5) {4};

\node[community2_node] (c2n1) at (4,2.5) {5};
\node[community2_node] (c2n2) at (5,3.2) {6};
\node[community2_node] (c2n3) at (5,1.8) {7};
\node[community2_node] (c2n4) at (6,2.5) {8};

\node[community3_node] (c3n1) at (2,0) {9};
\node[community3_node] (c3n2) at (3,0.7) {10};
\node[community3_node] (c3n3) at (3,-0.7) {11};
\node[community3_node] (c3n4) at (4,0) {12};

\graph[edges={edge_style}] {
    (c1n1) -- (c1n2);
    (c1n1) -- (c1n3);
    (c1n1) -- (c1n4);
    (c1n2) -- (c1n3); 
    (c1n2) -- (c1n4);
    (c1n3) -- (c1n4);
};

\graph[edges={edge_style}] {
    (c2n1) -- (c2n2);
    (c2n1) -- (c2n3);
    (c2n1) -- (c2n4);
    (c2n2) -- (c2n3); 
    (c2n2) -- (c2n4);
    (c2n3) -- (c2n4);
};

\graph[edges={edge_style}] {
    (c3n1) -- (c3n2);
    (c3n1) -- (c3n3);
    (c3n1) -- (c3n4);
    (c3n2) -- (c3n3); 
    (c3n2) -- (c3n4);
    (c3n3) -- (c3n4);
};

\path[edge_style] (c1n4) edge (c2n1); 
\path[edge_style] (c1n4) edge (c3n2); 
\path[edge_style] (c2n1) edge (c3n2); 


\end{tikzpicture}
}
}
\caption{(a) A simple network with some nodes that shows aspects of a community. (b) The same network, but with distinct communities. Nodes within each community (blue, green, and red) are more densely connected internally, with sparser connections between communities.}\label{fig:example_graph}
\end{figure}

Consider the toy network $G(V,E)$ shown in \autoref{fig:example_graph}(a). The graph representation converts a \texttt{NetworkX} or \texttt{igraph} network into an optimized data structure that maintains multiple complementary representations of the same graph data: (i) An edge list for sequential access. (ii) An adjacency list for neighborhood queries. (iii) Pre-computed degree mappings. (iv) An edge lookup hash set that enables constant-time edge existence queries. This design allows significant performance gains, providing $O(1)$ access times for operations.

An individual in the population is a candidate solution, represented as a hash map that maps each node $v \in V$ to a community identifier, as a mapping $\mathcal{M}$.
\[
  \mathcal{M}: V \;\longrightarrow\; \{c_1, c_2\dots\, c_k\}
\]
\noindent
For example of the \autoref{fig:example_graph}(a), consider the one possible mapping 
$\mathcal{M}=\{\,1\!:\!A,\;2\!:\!A,\;3\!:\!A,\;4\!:\!A,\;5\!:\!B,\;6\!:\!B,\;7\!:\!B,\;8\!:\!B,\;9\!:\!C,\;10\!:\!C,\;11\!:\!C,\;12\!:\!C\}$ showed in 
\autoref{fig:example_graph}(b).

This representation is memory-efficient, allowing for fast lookups. Our custom crossover and mutation operators drive the evolution of these individuals.

\subsection{Genetic Operators and Offspring Generation}
\label{sec:operators}
\subsubsection{Crossover}
\label{sec:ga-crossover}

The crossover operator is responsible for combining multiple parent solutions to generate a new offspring. Let $P = (P_1, \ldots, P_{N_p})$ denote a collection of ${N_p}$ parent individuals, where each individual $P_i$ represents a partition of the graph $G = (V, E)$, that is, a mapping from each node $v \in V$ to a community label $c \in C$. The operator is controlled by a crossover threshold parameter $C_R \in [0,1]$. Initially, a random number $x$ is sampled uniformly from the interval $[0,1]$. If $x > C_R$, the crossover is skipped, and one of the parent partitions is selected uniformly at random to serve as the offspring: 
\begin{equation}
  x > C_R \quad \Rightarrow \quad P_{\text{child}} = \text{Random}(P_1, \ldots, P_{N_p}).
\end{equation}

If $x \leq C_R$, the offspring is constructed by aggregating community assignments from all parent individuals. The process starts by considering the common node set $V$, and initializing an empty offspring mapping $P_{\text{child}}$. For each node $v \in V$, the algorithm counts the number of times each community label $c \in C$ has been assigned to $v$ across all parent individuals, using the following expression:
\begin{equation}
\text{count}(c, v) = \sum_{i=1}^{{N_p}} \mathds{1}\{P_i(v) = c\},
\end{equation}
where $\mathds{1}\{\cdot\}$ denotes the indicator function. The community (or communities) with the highest count is then identified as 
\begin{equation}
\label{eq:arg_max}
c^*(v) \in \arg\max_{c \in C} \text{count}(c, v).
\end{equation}
In the event of a tie, one of the top-scoring communities is selected uniformly at random. The selected label $c^*(v)$ is then assigned to node $v$ in the offspring partition, that is, $P_{\text{child}}(v) = c^*(v)$. This procedure is applied to all nodes in $V$, producing a complete offspring partition. The resulting solution tends to preserve dominant structural features from the parents while still allowing for variability and exploration through random tie-breaking. The crossover process is invoked for each of the ${N_p}$ offspring to be created in the new population, as detailed in the loop of Algorithm \ref{alg:hp-mocd-main} (lines 19-24), thus ensuring diversity and continuous exploration of the search space. 

\begin{algorithm}[!ht]
\small
\caption{Crossover Operator}\label{alg:crossover}
\begin{algorithmic}[1]
\Procedure{Crossover}{$P_1, \ldots, P_{N_p}$, $C_R$}
    \State \textbf{Input:} Collection of $N_p$ parent individuals $P = (P_1, \ldots, P_{N_p})$, crossover threshold $C_R \in [0,1]$
    \State \textbf{Output:} Offspring partition $P_{\text{child}}$
    
    \State Sample $x \sim \text{Uniform}[0,1]$
    \If{$x > C_R$}
        \State \Return $\text{Random}(P_1, \ldots, P_{N_p})$ \Comment{Skip crossover, return random parent}
    \EndIf
    
    \State Initialize $P_{\text{child}} \gets \emptyset$ \Comment{Empty offspring mapping}
    
    \State \Comment{\texttt{Aggregate community assignments from all parents}}
    \For{each node $v \in V$}
        \State \Comment{Count occurrences of each community label for node $v$}
        \For{each community $c \in C$}
            \State $\text{count}(c, v) \gets \sum_{i=1}^{N_p} \mathds{1}\{P_i(v) = c\}$
        \EndFor
        
        \State \Comment{Find community(ies) with highest count}
        \State $c^*(v) \gets \arg\max_{c \in C} \text{count}(c, v)$ \Comment{Random tie-breaking if multiple maximizers}
        \State $P_{\text{child}}(v) \gets c^*(v)$ \Comment{Assign selected community to node $v$}
    \EndFor
    \State \Return $P_{\text{child}}$
\EndProcedure
\end{algorithmic}
\end{algorithm}

\subsubsection{Mutation}
\label{sec:mutation}
To introduce variation and avoid premature convergence, HP-MOCD applies a mutation operator after crossover. This operator modifies the community assignment of selected nodes based on the structure of their local neighborhood. The mutation operator uses the same individual representation as described in Section~\ref{sec:ga-crossover}, along with a mutation threshold parameter \( M_R \in [0,1] \). Each node \( v \in V \) is considered for mutation independently: a random number \( r_v \in [0,1] \) is sampled, and nodes satisfying \( r_v < M_R \) are selected for mutation. The set of selected nodes is defined as:
\begin{equation}
V_m = \{ v \in V \mid r_v < M_R \}.
\end{equation}

For each selected node, the frequency of community assignments among its neighboring nodes is computed. To improve performance, the nodes are processed in batches. For a given node $v$, its set of neighbors $N(v)$ is retrieved, and for each neighbor $u \in N(v)$, the community to which $u$ belongs is noted. The algorithm then determines the most frequent community using the same \autoref{eq:arg_max}. 

\subsection{Solution Selection}

For the multi-objective optimization problem formulated in this study, the goal is to find solutions that
``minimize'' $\mathbf{f}(C)$ (see \autoref{eq:problem}) in the Pareto-optimality context. Algorithms such as NSGA-II do not return a single ``optimal'' solution. Instead, return a set of feasible, mutually non-dominated solutions that approximate as much as possible the Pareto set, which represents the optimal trade-off among the multiple objectives \cite{moreira:19}.

The Decision Maker (DM) must choose the most suitable option from the approximation set to ``implement''.

Following the strategy proposed by \cite{shi2010}, we adopt a scalarization-based criterion to identify the solution that maximizes the following quality function:
\begin{equation}
\label{eq:selection_method_equation}
Q(C) = 1 - f_1 - f_2
\end{equation}

\begin{figure}[!th]
  \centering
    \subfigure[Pareto Frontier]{
      \includegraphics[width=0.45\textwidth]{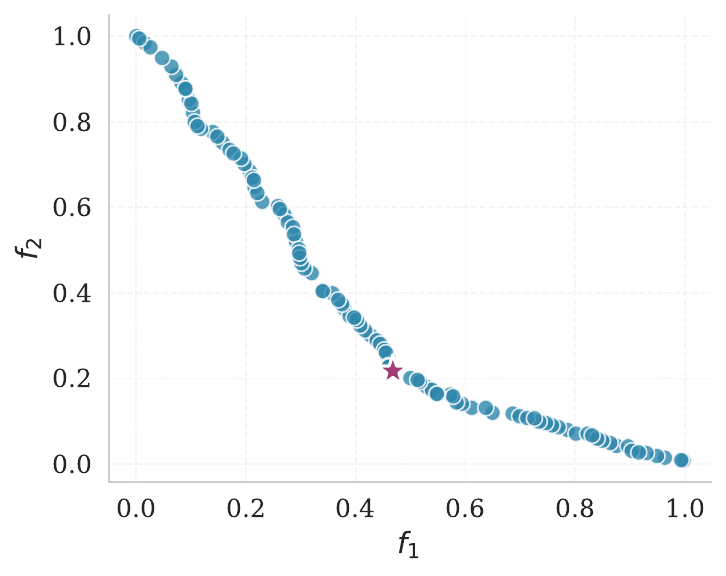} 
    }
    \subfigure[$Q(C)$ vs. Number of Communities]{
      \includegraphics[width=0.45\textwidth]{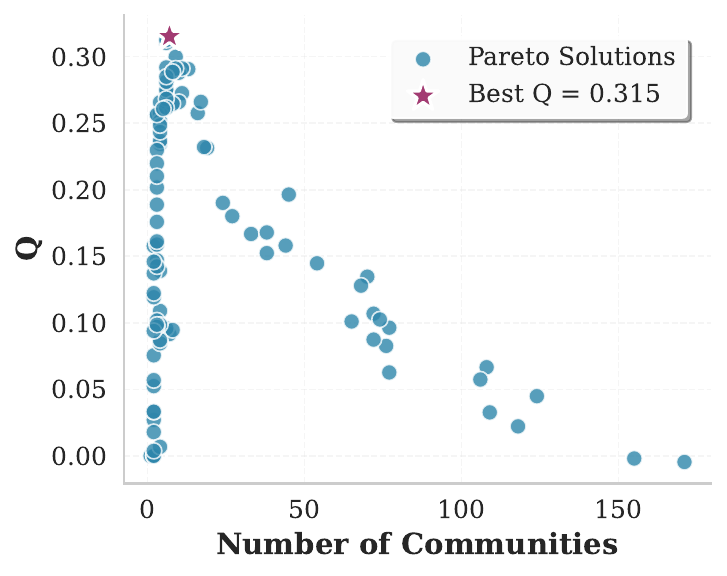}
    }
    \subfigure[$Q(C)$ vs. NMI]{
      \includegraphics[width=0.45\textwidth]{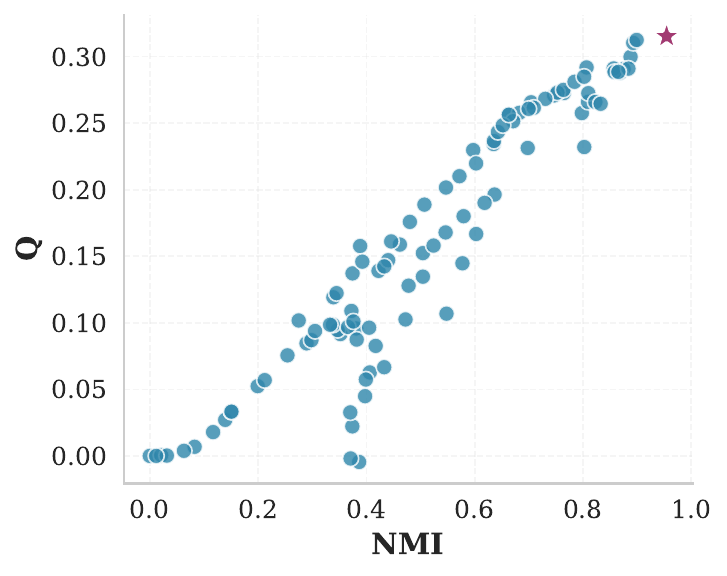}
    }
    \subfigure[$Q(C)$ vs. AMI]{
      \includegraphics[width=0.45\textwidth]{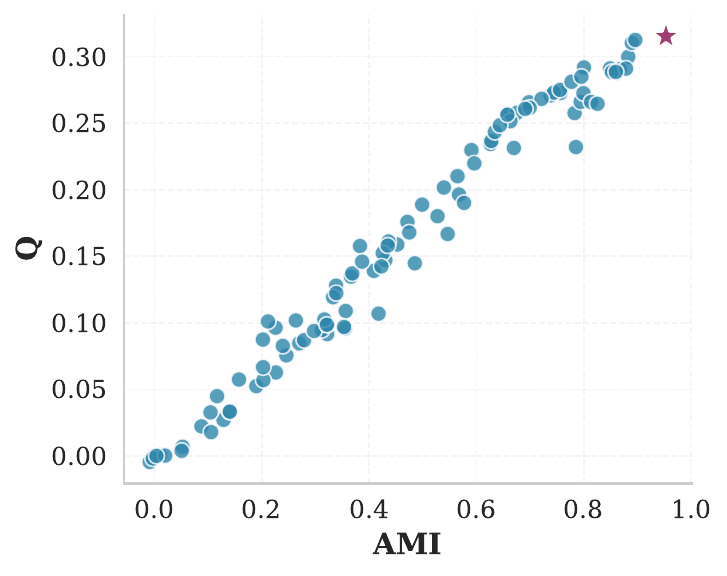}
    }
    \caption{Analysis of the scalarization-based selection method. The red dot highlights the solution with the highest value of $Q(C)$.}
    \label{fig:pareto_frontier_selection_method_analisys}
\end{figure}

This score favors solutions with low intra-community penalty and low inter-community connectivity, as defined in \autoref{eq:objectives}, thus promoting partitions that are both cohesive and well-separated. \autoref{fig:pareto_frontier_selection_method_analisys} illustrates this behavior: the red dot indicates the solution that maximizes \( Q(C) \). We observe that this selection strategy consistently identifies partitions that also perform well under NMI and AMI validation metrics.

\begin{algorithm}[!ht]
\small
\caption{Solution Selection}\label{alg:selection}
\begin{algorithmic}[1]
\Procedure{SelectSolution}{Pareto front $\mathcal{F}_1$}
    \State \textbf{Input:} Set of non-dominated solutions $\mathcal{F}_1 = \{p_1, p_2, \ldots, p_k\}$
    \State \textbf{Output:} Recommended solution $p^*$
    
    \State \Comment{\texttt{Scalarization-based selection criterion}}
    \State $\text{selected\_solution} \gets -\infty$
    \State $p^* \gets \text{null}$
    
    \For{each solution $p \in \mathcal{F}_1$}
        \State $Q(p) \gets 1 - f_1(p) - f_2(p)$ 
        \If{$Q(p) > \text{selected\_solution}$}
            \State $\text{selected\_solution} \gets Q(p)$
            \State $p^* \gets p$
        \EndIf
    \EndFor
    
    \State \Comment{\texttt{Domain-specific selection criteria can be applied here}}
    \State \Comment{\texttt{e.g., number of key nodes, communities, conductance, etc.}}
    
    \State \Return $p^*$
\EndProcedure
\end{algorithmic}
\end{algorithm}

An important advantage of the Pareto-based approach is the flexibility it affords in selecting solutions based on application-specific preferences. While we adopt the scalarization function \( Q(C) \) as a default selection criterion, practitioners are not constrained to this choice. Depending on the context, alternative metrics can be employed to guide selection from the Pareto front. For instance, one might prioritize solutions based on modularity, the number of detected communities, average community size, or even the variance in community sizes. Other structural properties, such as conductance, cut ratios, or expansion scores, may also be relevant in certain domains. This flexibility allows users of HP-MOCD to serve as a general-purpose tool, as it provides not just one partition, but a rich set of candidate solutions from which the most appropriate can be chosen using auxiliary metrics or domain-specific constraints. This is a fundamental departure from single-objective methods, which collapse the search space into a single score and, therefore, offer limited interpretability and adaptability.

\subsection{Complexity Analysis}

Complexity time of HP-MOCD per generation is driven by its genetic operators (crossover and mutation) and the key NSGA-II routines (non-dominated sorting and crowding distance). Let $|V|$ be the number of nodes, $|E|$ the number of edges, $N_p$ the population size, and $M_R$ the mutation rate.

\subsubsection{Operators Complexity Analysis}

\paragraph{Crossover \& Mutation}
For each node in the graph, the algorithm crossover identifies the most frequent community label assigned to that node among a fixed number of parent solutions. Since hash map operations have an expected constant time complexity, $O(1)$, processing each node takes $O(1)$ time. Assigning the most frequent label also takes constant time per node. Consequently, for all $|V|$ nodes, the expected time complexity of a single crossover operation is linear in the number of nodes, i.e., $O(|V|)$.

The mutation iterates through each of the $|V|$ nodes, selecting a node for mutation with a given probability $M_R \in [0,1]$. This selection phase involves a random draw for each node, costing $O(|V|)$ time, and, for each node selected for mutation, inspects its local neighborhood to determine a new community assignment (by finding the most frequent community label among its neighbors). This requires visiting all adjacent nodes, so the cost for mutating a single node $v$ is proportional to its degree, $d(v)$.
The expected number of nodes selected for mutation is $M_R \cdot |V|$. The total time for recalculating assignments for these mutated nodes is proportional to the sum of their degrees. In expectation, this sum is $M_R \sum_{v \in V} d(v) = M_R \cdot 2|E|$. Therefore, the expected time complexity for this part of the mutation is $O(M_R \cdot |E|)$. Combining the node selection phase and the assignment update phase, the total time complexity for one mutation operation is $O(|V| + M_R \cdot |E|)$. For sparse graphs, where $|E| = O(|V|)$, and assuming $M_R$ is a constant, this simplifies to $O(|V|)$.

\paragraph{Non-Dominated Sorting \& Crowding Distance Calculation}
These operators act upon a combined population of $2N_p$ individuals (current population $P_t$ and offspring population $Q_t$, each of size $N_p$). This process sorts the $2N_p$ individuals into different non-domination fronts. The non-dominated sorting of $N$ individuals can be performed in $O(N \log N)$ time with fixed 2 objectives. Thus, for the $2N_p$ individuals, this step takes $O((2N_p) \log (2N_p)) = O(N_p \log N_p)$ time. 

For each front of size $F_i$, solutions are sorted based on each of the $M$ objective functions. This takes $O(M F_i \log F_i)$ time per front. Summing over all fronts $\sum F_i = 2N_p$, the total time complexity for crowding distance calculation across all fronts is $O(M (2N_p) \log (2N_p))$. For $M=2$, this becomes $O(N_p \log N_p)$. 

\paragraph{Overall Complexity}

The dominant cost per generation is the sum of: $T_{\text{gen}} = O(N_p |V|) + O(N_p \log N_p) + O(N_p \log N_p)$ which simplifies to $T_{\text{gen}} = O(N_p |V| + N_p \log N_p)$. Since $|V|$ in $G$ is typically much larger than $\log N_p$, the $O(N_p |V|)$ term usually dominates. Thus, the complexity per generation effectively becomes: $O(N_p |V|)$ for sparse graphs. Over $G$ generations, the total time complexity of HP-MOCD is: $O(G \cdot N_p |V|)$. 
A detailed analysis of the algorithm's space complexity, including the dominant memory terms across components, is provided in \autoref{sec:appendixA}.

\section{Experimental Results}
\label{sec:experiments}
This section details the methodology and experimental evaluation of HP-MOCD. Our study explores four core dimensions: (i) detection performance on synthetic and real-world networks, (ii) scalability with increasing network size, (iii) parallel efficiency via multithreaded execution, and (iv) robustness under structural noise. Experiments are conducted using standard benchmarks, well-established metrics, and strong baseline algorithms, ensuring fair and reproducible comparisons. To ensure rigorous and interpretable comparisons, we perform pairwise statistical tests (paired t-test with initial $\alpha = 0.05$), applying Bonferroni correction to adjust for multiple comparisons. The goal is to determine, for each evaluation, whether HP-MOCD is statistically superior, equivalent, or inferior to the best-performing baseline. We classify each result accordingly and highlight them in \textbf{bold} (superior) in the tables. 

\subsection{Comparison Algorithms and Parameter Configuration}
\label{sec:baselines}

We compare HP-MOCD against a diverse set of classical heuristics and multi-objective evolutionary algorithms, described below:

\begin{itemize}
    \item \textbf{MOCD}~\cite{shi2010}: A two-objective evolutionary method (based on PESA-II) optimizing intra-community density and inter-community sparsity.
    \item \textbf{MOGA‑Net}~\cite{Pizzuti2009}: A modularity-oriented genetic algorithm evolving partitions directly.
    \item \textbf{Louvain}~\cite{blondel2008}: A multilevel greedy heuristic that locally optimizes modularity.
    \item \textbf{Leiden}~\cite{traag2019}: A refinement of Louvain ensuring connected communities and improved convergence.
    \item \textbf{ASYN-LPA}~\cite{raghavan2007near}: A fast, parameter-free method based on label propagation among neighbors.
    \item \textbf{KRM}~\cite{shaik2021evolutionary}: An NSGA-III variant optimizing Kernel-k-means, Ratio-cut, and Modularity.
    \item \textbf{CCM}~\cite{shaik2021evolutionary}: An NSGA-III variant targeting Community Fitness, Community Score, and Modularity.
    \item \textbf{CDRME}~\cite{dabaghi2025community}: A decomposition-based evolutionary algorithm using spectral-based seeding and optimizing multiple modularity-related objectives.
\end{itemize}

All algorithms were implemented in Python, except HP-MOCD, which was implemented in Rust and exposed to Python via PyO3\footnote{\url{https://pyo3.rs/v0.25.0/}}. Although HP-MOCD exploits parallelism in population-level operators, runtime comparisons reported in Sections~\ref{sec:robustness} and~\ref{sec:scalability} were performed in single-threaded mode except for \methodname. Single-thread and parallel performance are evaluated separately in \autoref{sec:appendixB}. The algorithm baseline configurations are summarized in \autoref{tab:algorithms_setup}, based on their original publications.

\begin{figure}[!th]
  \centering
    \subfigure[$maxgen = 5$]{
      \includegraphics[width=0.30\textwidth]{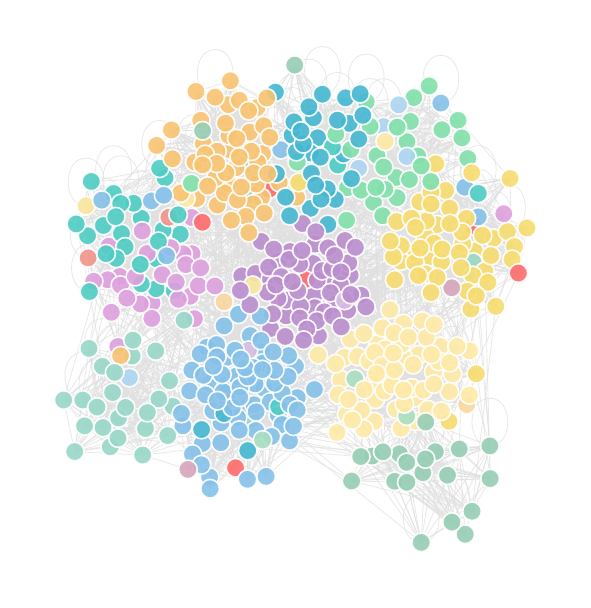} 
    }
    \subfigure[$maxgen = 50$]{
      \includegraphics[width=0.30\textwidth]{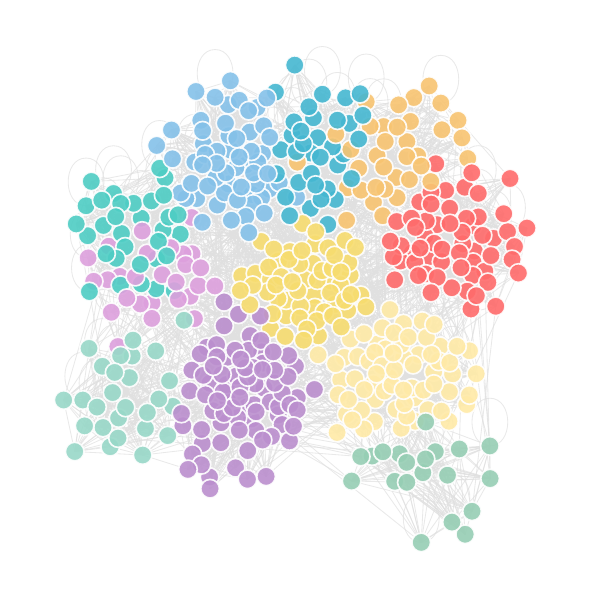} 
    }
    \subfigure[$maxgen$ = 100]{
      \includegraphics[width=0.30\textwidth]{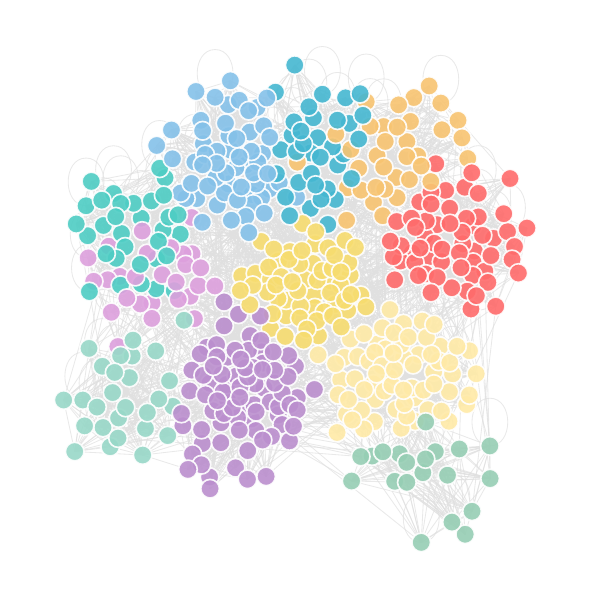} 
    }
    \caption{Performance results across varying $maxgen$ values with $\mu = 0.1$. Refer to \autoref{sec:paramters-setup} for details on the experimental setup.}
    \label{fig:comparacao-nsga-maxgen}
\end{figure}

The evolutionary process in HP-MOCD is governed by two key hyperparameters: population size and number of generations. To determine appropriate values, we conducted empirical experiments on synthetic benchmarks, using the same configuration as in our main experiments, except for varying $n$ (see Section~\ref{sec:synthetic_benchmark_networks}). We analyzed the convergence behavior of the algorithm under these settings. As shown in \autoref{fig:comparacao-nsga-maxgen}, 100 generations provided a favorable trade-off between solution quality and computational cost; for the other parameters and scenarios, we leave them in the \autoref{sec:appendixC}. HP-MOCD used the following configuration: crossover probability \( C_P = 0.8 \), mutation probability \( M_P = 0.2 \), and number of parents per crossover \( E_S = 4 \). \autoref{tab:algorithms_setup} summarizes the configurations.

\begin{table}[!ht]
    \centering
    \caption{Parameter settings and types of the compared algorithms. ``Gens" and ``Pop" denote the number of generations and population size for evolutionary algorithms. $C_P$, $M_P$, and $E_S$ refer to crossover probability, mutation probability, and number of parents per crossover, respectively.}
    \label{tab:algorithms_setup}
    \begin{tabular}{llcccccc}
        \toprule
        \textbf{Algorithm} & \textbf{Type} & \textbf{Gens} & \textbf{Pop} & $C_P$ & $M_P$ & $E_S$ & \textbf{Reference} \\
        \midrule
        ASYN-LPA   & Label Prop. & --   & --   & --  & --  & -- & \cite{raghavan2007near} \\
        Louvain    & Greedy       & --   & --   & --  & --  & -- & \cite{blondel2008} \\
        Leiden     & Greedy       & --   & --   & --  & --  & -- & \cite{traag2019} \\
        MOGA‑Net   & MOEA         & 100  & 100  & 0.9 & 0.1 & -- & \cite{Pizzuti2009} \\
        MOCD       & MOEA         & 100  & 100  & 0.9 & 0.1 & -- & \cite{shi2010} \\
        KRM        & MOEA         & 200  & 100  & 0.9 & 0.1 & -- & \cite{shaik2021evolutionary} \\
        CCM        & MOEA         & 200  & 100  & 0.9 & 0.1 & -- & \cite{shaik2021evolutionary} \\
        CDRME      & MOEA         & 100  & 100  & 0.9 & 0.2 & -- & \cite{dabaghi2025community} \\
        HP‑MOCD    & MOEA         & 100  & 100  & 0.8 & 0.2 & 4  & Ours \\
        \bottomrule
    \end{tabular}
\end{table}

\subsubsection{Evaluation Metrics}

We assess the similarity between the detected and ground-truth community structures using three widely adopted external validation metrics: Normalized Mutual Information (NMI), Adjusted Mutual Information (AMI), and the pairwise F1-score \cite{Vinh:2009, Rossetti2019CDLib}. These metrics evaluate how well the detected partition recovers the original grouping of nodes, from both an information-theoretic and structural perspective. Throughout our experiments, these metrics are used in a complementary fashion to capture different aspects of clustering quality.

Unlike standard classification tasks, community detection does not assume fixed label identities—only the grouping of nodes into communities matters. As such, traditional per-label precision and recall-based F1-scores are not applicable. Instead, we adopt a pairwise evaluation framework, in which all unordered node pairs are treated as binary decisions such that each pair is considered positive if both nodes belong to the same community, and negative otherwise. Precision and recall are computed by comparing these pairwise co-assignment decisions in the predicted and ground-truth partitions, and the F1-score is their harmonic mean \cite{Vinh:2009}. Each metric captures a different aspect of clustering quality. We use them jointly to ensure a comprehensive assessment. Below, we describe each of these metrics in detail.

\begin{itemize}
    \item \textbf{Normalized Mutual Information (NMI):} This metric quantifies the agreement between the detected community partition \( C \) and the ground-truth partition \( C' \) \cite{Yao2003-NMI-Original, vinh2010bailey-NMI-Sklearn}, defined as:
\begin{equation}
\text{NMI}(C, C') = \frac{-2 \sum_{i=1}^{n_1} \sum_{j=1}^{n_2} M_{ij} \log\left(\frac{M_{ij} N}{M_{i.} M_{.j}}\right)}{\sum_{i=1}^{n_1} M_{i.} \log\left(\frac{M_{i.}}{N}\right) + \sum_{j=1}^{n_2} M_{.j} \log\left(\frac{M_{.j}}{N}\right)},
\end{equation}
where \( M_{ij} \) is the confusion matrix between partitions, \( N \) is the number of nodes, and \( M_{i.} \), \( M_{.j} \) are marginal sums. NMI ranges from 0 (no agreement) to 1 (perfect match).

    \item \textbf{Adjusted Mutual Information (AMI):} AMI corrects the mutual information for chance \cite{vinh2010bailey-NMI-Sklearn}, offering greater robustness when the number or size of communities varies:
\begin{equation}
\text{AMI}(U, V) = \frac{\text{MI}(U, V) - \mathbb{E}[\text{MI}(U, V)]}{\max(H(U), H(V)) - \mathbb{E}[\text{MI}(U, V)]},
\end{equation}
where \( \text{MI}(U, V) \) is the mutual information between partitions \( U \) and \( V \), \( \mathbb{E}[\text{MI}] \) is its expected value under random labeling, and \( H(\cdot) \) denotes entropy. Like NMI, AMI ranges from 0 to 1 but is less sensitive to random or imbalanced partitions.
    \item \textbf{F1-Score:} This metric evaluates clustering accuracy by analyzing all unordered node pairs. Each pair is treated as a binary classification, where it is labeled positive if both nodes belong to the same community, and negative otherwise. Comparing these assignments between the predicted and ground-truth partitions, we compute \cite{manning2008ir, Vinh:2009}:
\[
\text{F1} = 2 \cdot \frac{\text{Precision} \cdot \text{Recall}}{\text{Precision} + \text{Recall}},
\]
where precision is the fraction of predicted intra-community pairs that are also present in the ground truth, and recall is the fraction of true intra-community pairs correctly recovered. This pairwise evaluation captures the global consistency of group assignments and is invariant to label permutations.

\item \textbf{Modularity:} Although modularity is not an external validation metric, it remains widely used to assess the internal structure of communities based on edge density~\cite{newman2006modularity}. It compares the observed number of edges within communities to the expected number in a randomized null model that preserves the degree distribution. In essence, modularity quantifies how much more densely connected the nodes are within communities than would be expected by chance. Its score theoretically ranges from \(-\frac{1}{2}\) to 1, with higher values indicating stronger community structure. It is formally defined as:
\begin{equation}\label{eq:modularity}
Q = \frac{1}{2m} \sum_{i,j} \left( A_{ij} - \frac{k_i k_j}{2m} \right) \delta(c_i, c_j)
\end{equation}
where \( A_{ij} \) is the adjacency matrix, \( k_i \) and \( k_j \) are the degrees of nodes \( i \) and \( j \), \( m \) is the total number of edges, and \( \delta(c_i, c_j) \) is 1 if nodes \( i \) and \( j \) are assigned to the same community and 0 otherwise. We include modularity in our evaluation to allow direct comparison with modularity-optimizing methods such as Louvain and Leiden.

\end{itemize}

\subsection{Synthetic Benchmark Networks}
\label{sec:synthetic-networks}
\label{sec:synthetic_benchmark_networks}
We use the LFR benchmark generator~\cite{lfr2008}, as implemented in the \texttt{networkx} Python library\footnote{\url{https://networkx.org/documentation/stable/reference/generated/networkx.generators.community.LFR_benchmark_graph.html}}, to generate synthetic networks with known ground-truth communities. LFR is widely adopted due to its ability to model realistic topological characteristics such as power-law degree distributions and heterogeneous community sizes. We adopt two main configurations:

\begin{itemize}
    \item \textbf{Robustness test:} Fixing network size \( n = 1{,}000 \), we vary the mixing parameter \( \mu \in [0.1, 0.8] \) in increments of 0.1. This parameter controls the fraction of links that connect each node to other communities, such that lower values imply well-separated clusters, while higher values introduce more structural ambiguity.

    \item \textbf{Scalability test:} Network size \( n \in \{500, 1{,}000, 1{,}500, 2{,}000, \ldots, 20{,}000\} \), with average degree 20 and maximum degree 50. Community sizes vary in \([20, 100]\), and other parameters are fixed as: degree exponent \( \tau_1 = 2.5 \), community size exponent \( \tau_2 = 1.5 \), no overlapping nodes, and one community per node.

\end{itemize}

All configurations were replicated with 20 different random seeds, and average scores with standard deviation are reported.

\subsubsection{Robustness to Community Mixing}\label{sec:robustness}

\begin{figure}[!th]
  \centering
    \subfigure[NMI]{
      \includegraphics[width=0.45\textwidth]{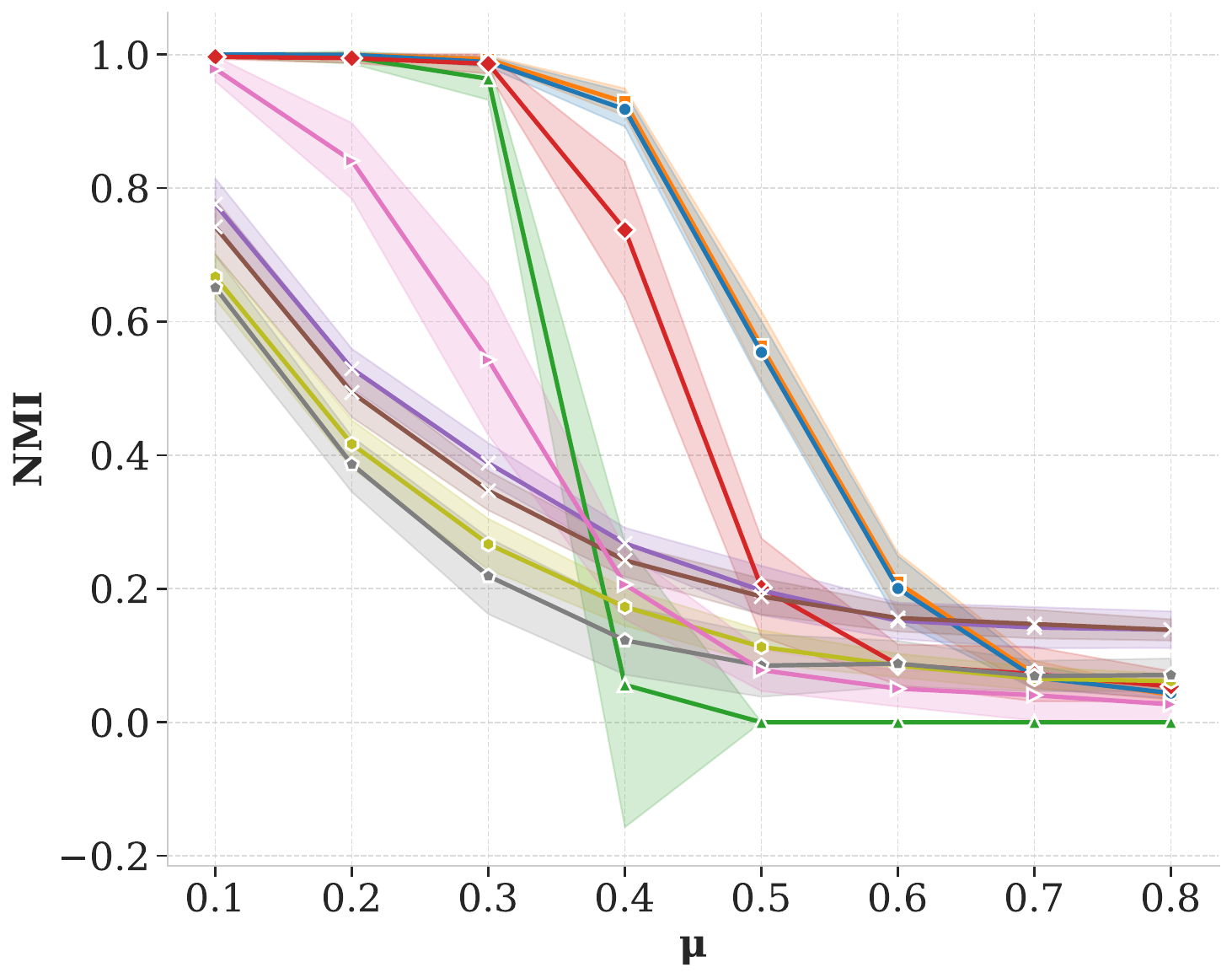} 
    }
    \subfigure[AMI]{
      \includegraphics[width=0.45\textwidth]{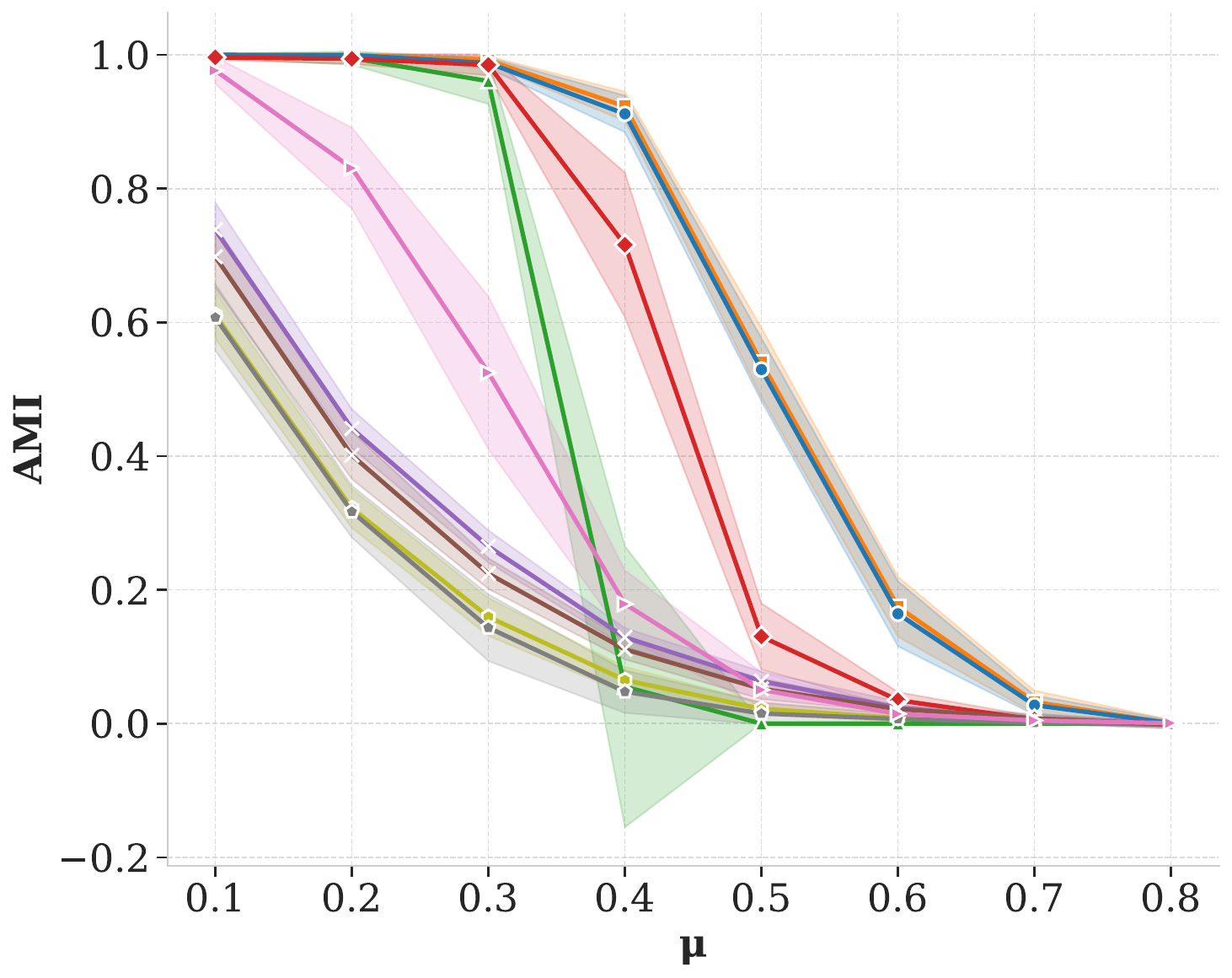} 
    }
    \subfigure[Modularity]{
      \includegraphics[width=0.45\textwidth]{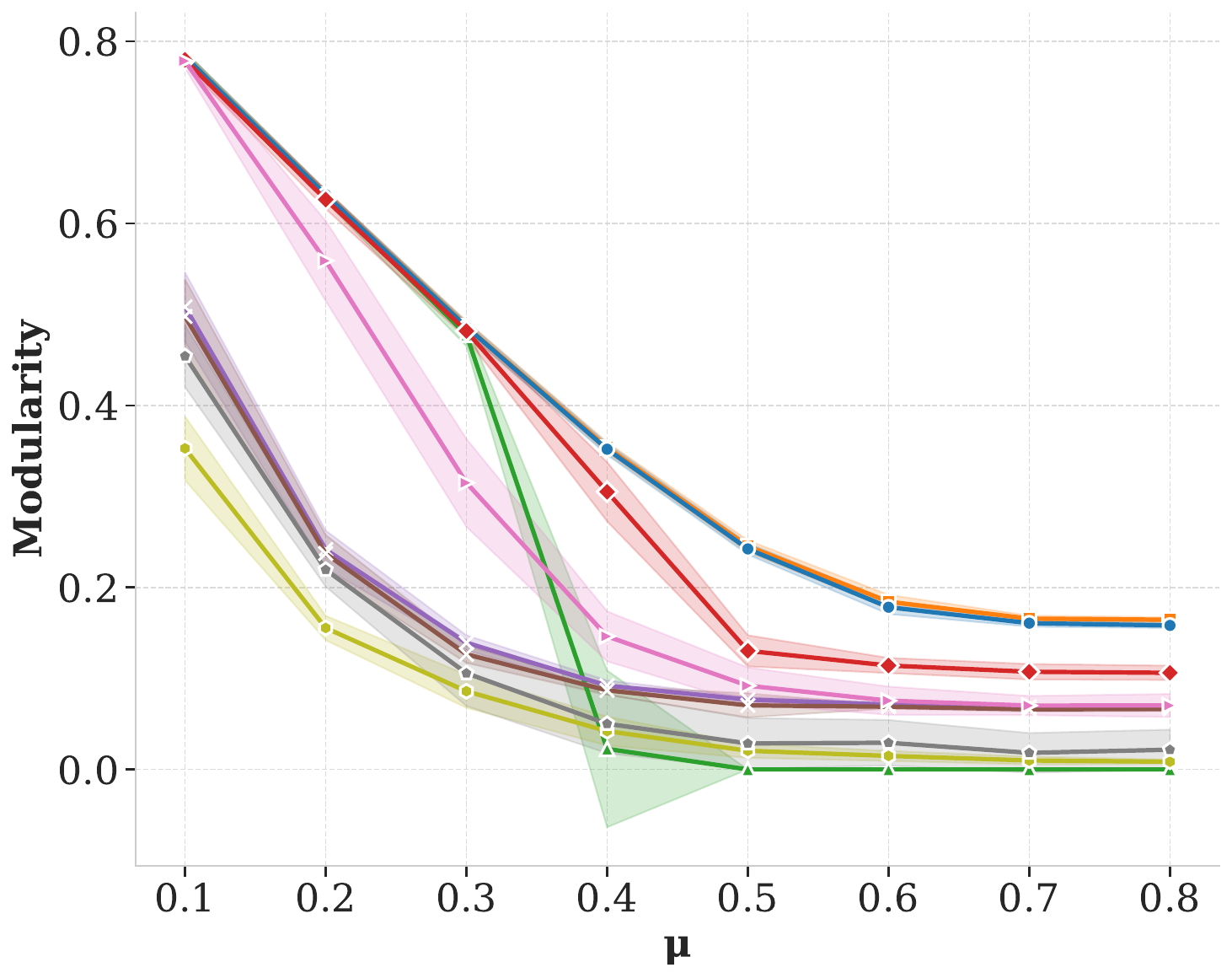} 
    }
    \subfigure[Execution Time (s)]{
      \includegraphics[width=0.45\textwidth]{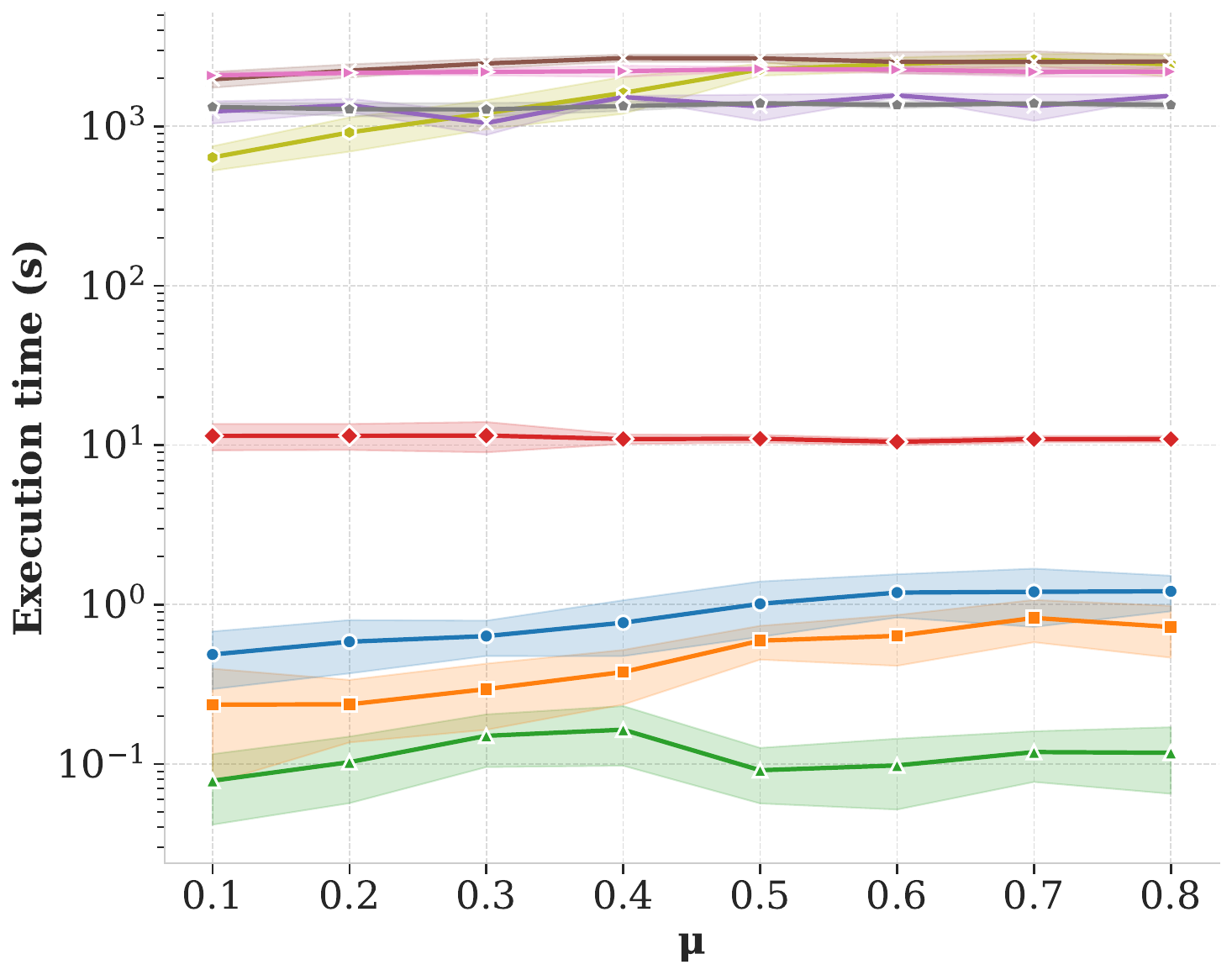} 
    }
    \subfigure{
      \includegraphics[width=0.7\textwidth]{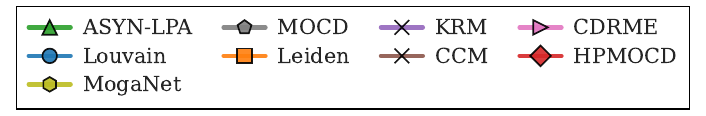} 
    }
    \caption{Performance of the evaluated algorithms on LFR networks with varying levels of structural noise ($\mu$), using 1{,}000 nodes. Each result represents the mean over 20 runs. Metrics shown include: (a) Normalized Mutual Information (NMI), (b) Adjusted Mutual Information (AMI), (c) modularity, and (d) execution time in seconds. Shaded areas indicate 95\% confidence intervals.}
    \label{fig:experiment-mu}
\end{figure}

\autoref{fig:experiment-mu} presents the performance of the algorithms under increasing levels of structural noise, controlled by the mixing parameter $\mu$ in LFR benchmark networks. This parameter determines the proportion of edges that connect nodes to different communities: lower $\mu$ values represent well-separated clusters, while higher values introduce greater inter-community connectivity, making the detection task more challenging.

All evaluations were conducted on synthetic networks with 1{,}000 nodes, using 20 independent seeds for each $\mu \in \{0.1, 0.2, \ldots, 0.8\}$. For each configuration, we report the mean values and 95\% confidence intervals of the evaluation metrics.

As shown in \autoref{fig:experiment-mu}(a) and (b), the proposed HP-MOCD algorithm achieves high detection accuracy across most values of $\mu$. Both NMI and AMI remain high and stable up to $\mu = 0.4$, indicating strong robustness to moderate levels of community mixing and structural noise. This suggests that HP-MOCD reliably maintains coherent partitions even when boundaries between communities become less distinct.

As expected, performance declines for all methods when $\mu \geq 0.4$, where community structure becomes increasingly ambiguous. However, HP-MOCD remains superior to most multi-objective baselines and continues to rival Louvain, Leiden, and ASYN-LPA in terms of NMI and AMI in this high-noise regime. These results emphasize HP-MOCD’s ability to adapt to challenging conditions while maintaining the advantages of its multi-objective design, such as generating a diverse set of near-optimal solutions.

The modularity trends shown in \autoref{fig:experiment-mu}(c) further highlight HP-MOCD’s effectiveness. It consistently achieves the highest modularity values among all MOEA-based methods, and even surpasses ASYN-LPA, a fast label propagation method sensitive to modularity. Despite not explicitly optimizing for modularity alone, HP-MOCD’s results remain comparable to Louvain and Leiden for $\mu \leq 0.3$, confirming its ability to uncover dense intra-community structures in well-separated scenarios.

Execution time comparisons in \autoref{fig:experiment-mu}(d) demonstrate that HP-MOCD is also computationally efficient. Unlike several MOEA-based methods that suffer from unstable or sharply increasing runtimes near $\mu = 0.6$ (in log scale), HP-MOCD maintains stable execution times across all levels of noise. For 15{,}500 nodes. \methodname~ was 181.4x faster than KRM, 259.4x than CCM. For 13{,}000 nodes 132.1x faster than MOCD and 553.1x faster than CDRME for 2{,}000 nodes. This robustness is attributed to its scaling only with the number of nodes in the network, its efficient parallel design, and linear-time genetic operators, as discussed in Section~\ref{sec:proposed_algorithm}.

In summary, in the synthetic experiments \autoref{fig:experiment-mu}, HP-MOCD achieves statistically the same performance in terms of NMI,  AMI and Modularity up to $\mu \leq 0.3$ ($p < 0.01$). At the extreme noise level ($\mu > 0.3$), all methods deteriorate. Thus, HP-MOCD demonstrates strong robustness to increasing structural noise, delivering accurate, modular, and scalable community detection.

\subsubsection{Scalability with Network Size}\label{sec:scalability}

\begin{figure}[t]
  \centering
    \subfigure[NMI]{
      \includegraphics[width=0.45\textwidth]{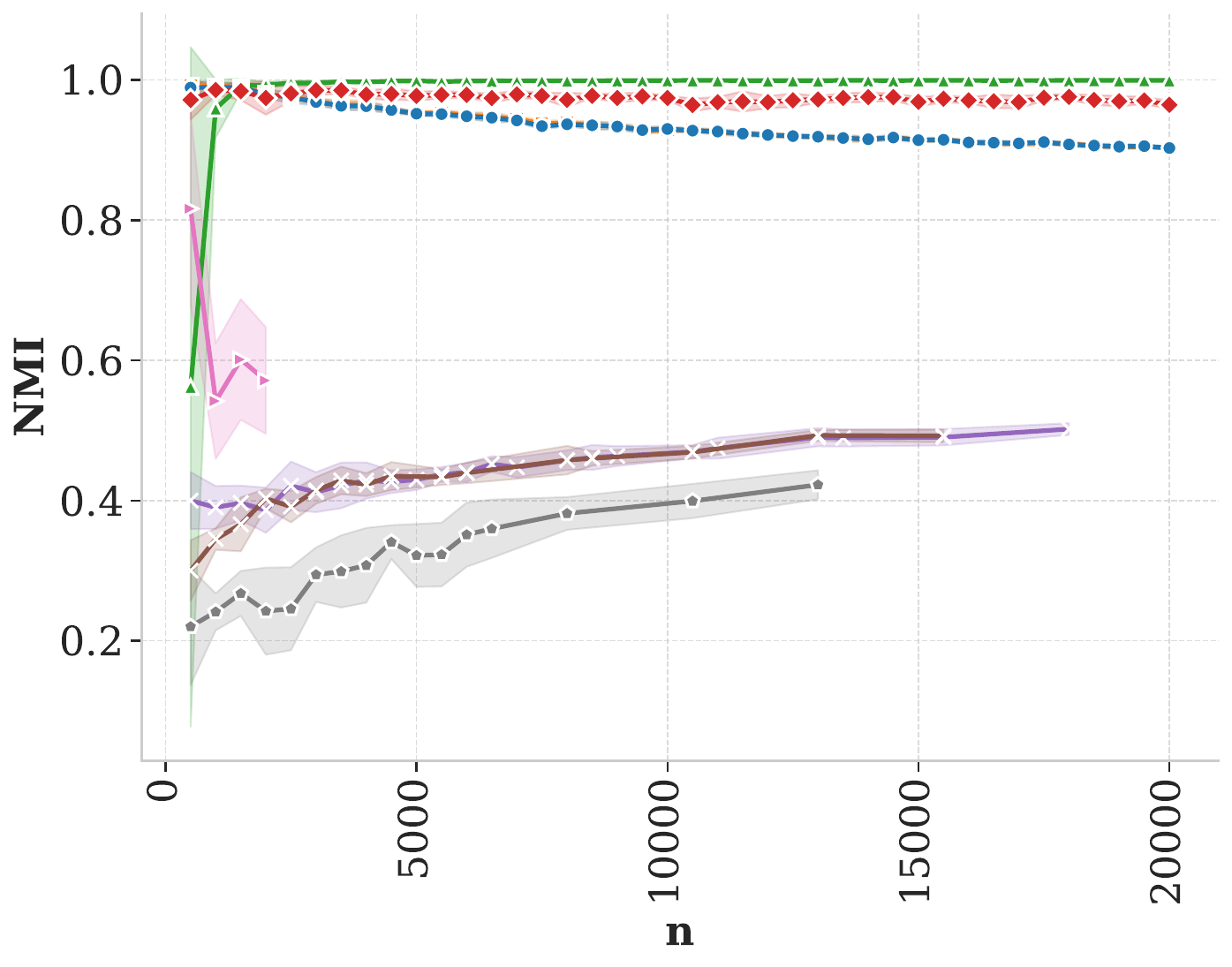} 
    }
    \subfigure[AMI]{
      \includegraphics[width=0.45\textwidth]{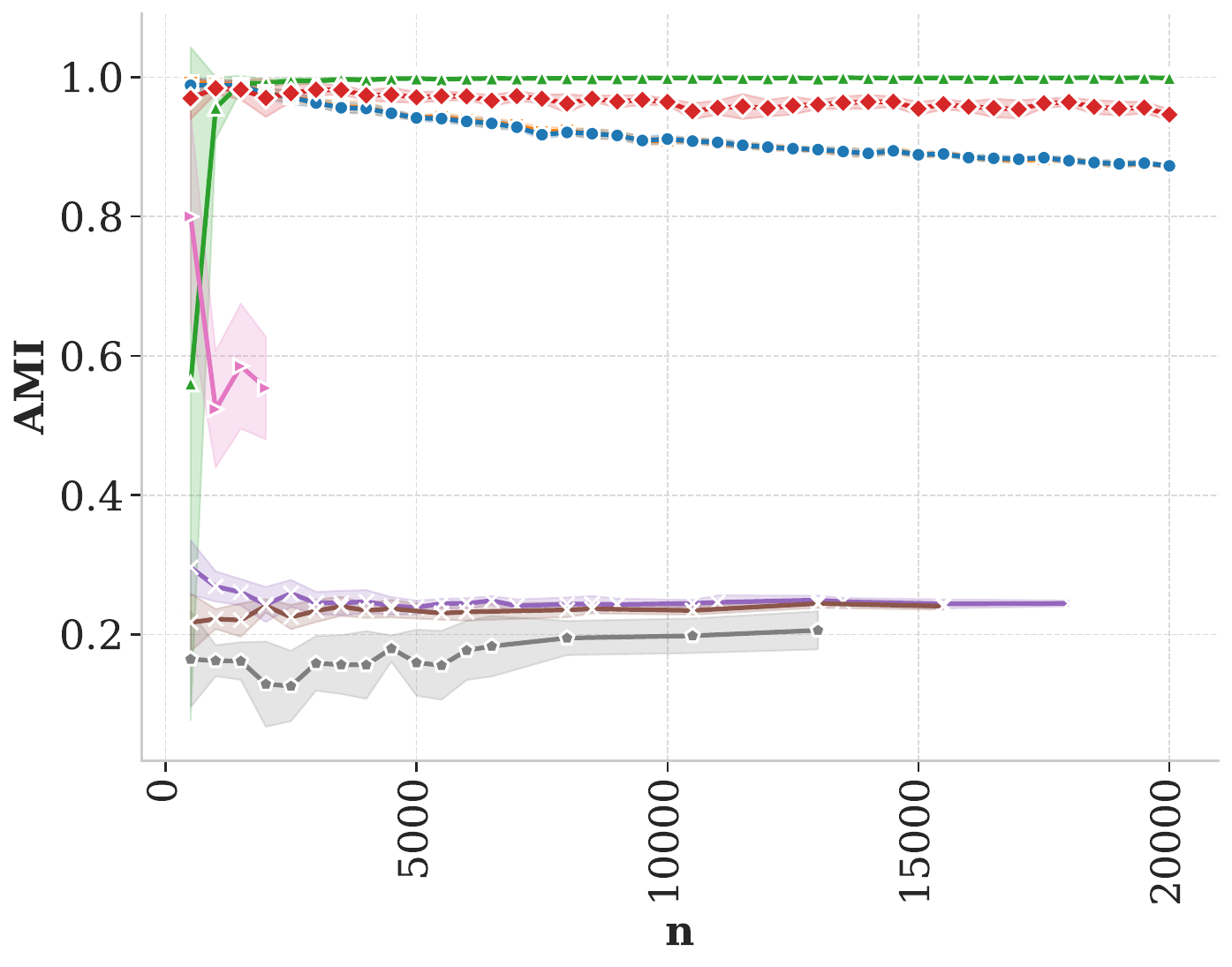} 
    }
    \subfigure[Modularity]{
      \includegraphics[width=0.45\textwidth]{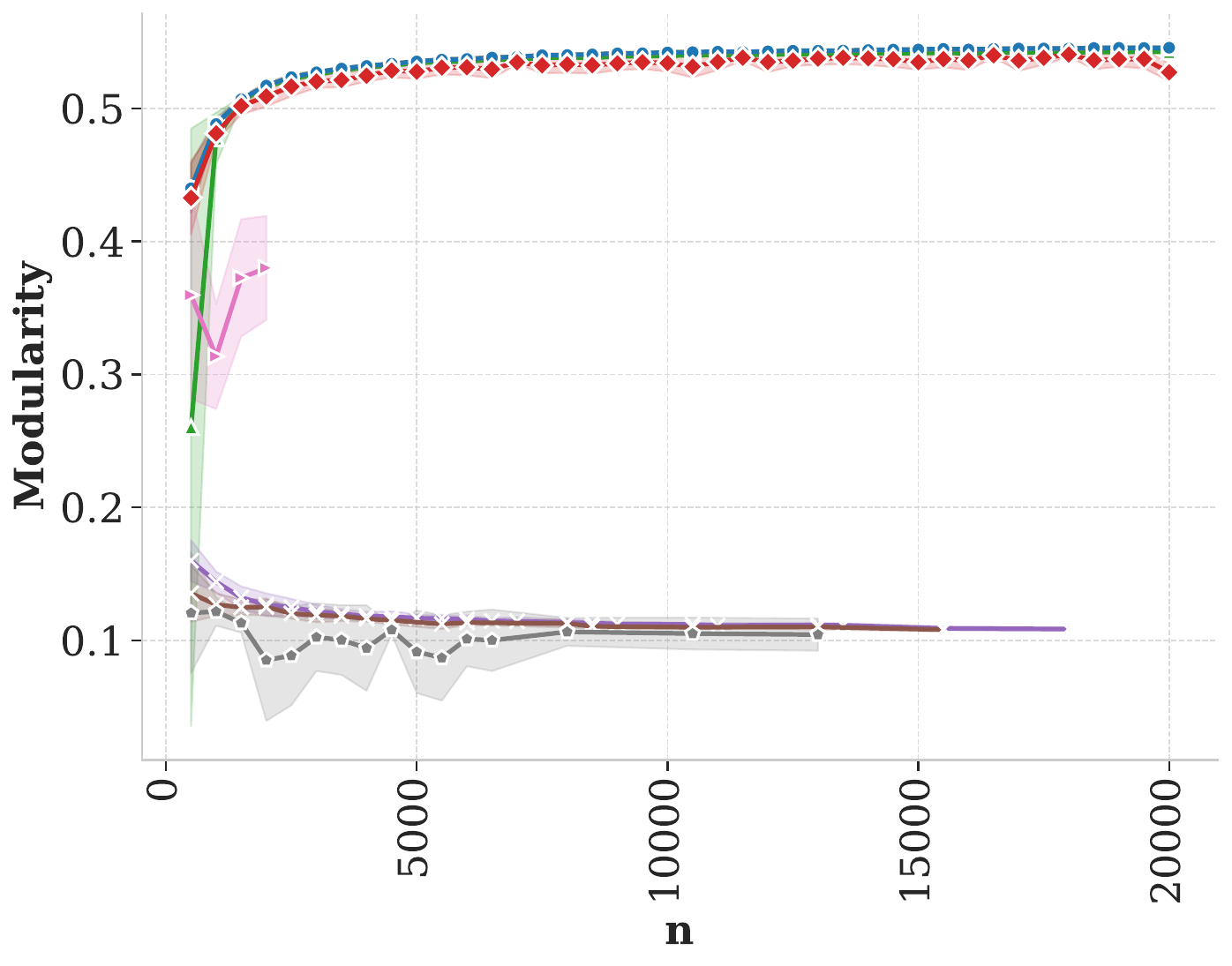} 
    }
    \subfigure[Execution Time (s)]{
      \includegraphics[width=0.45\textwidth]{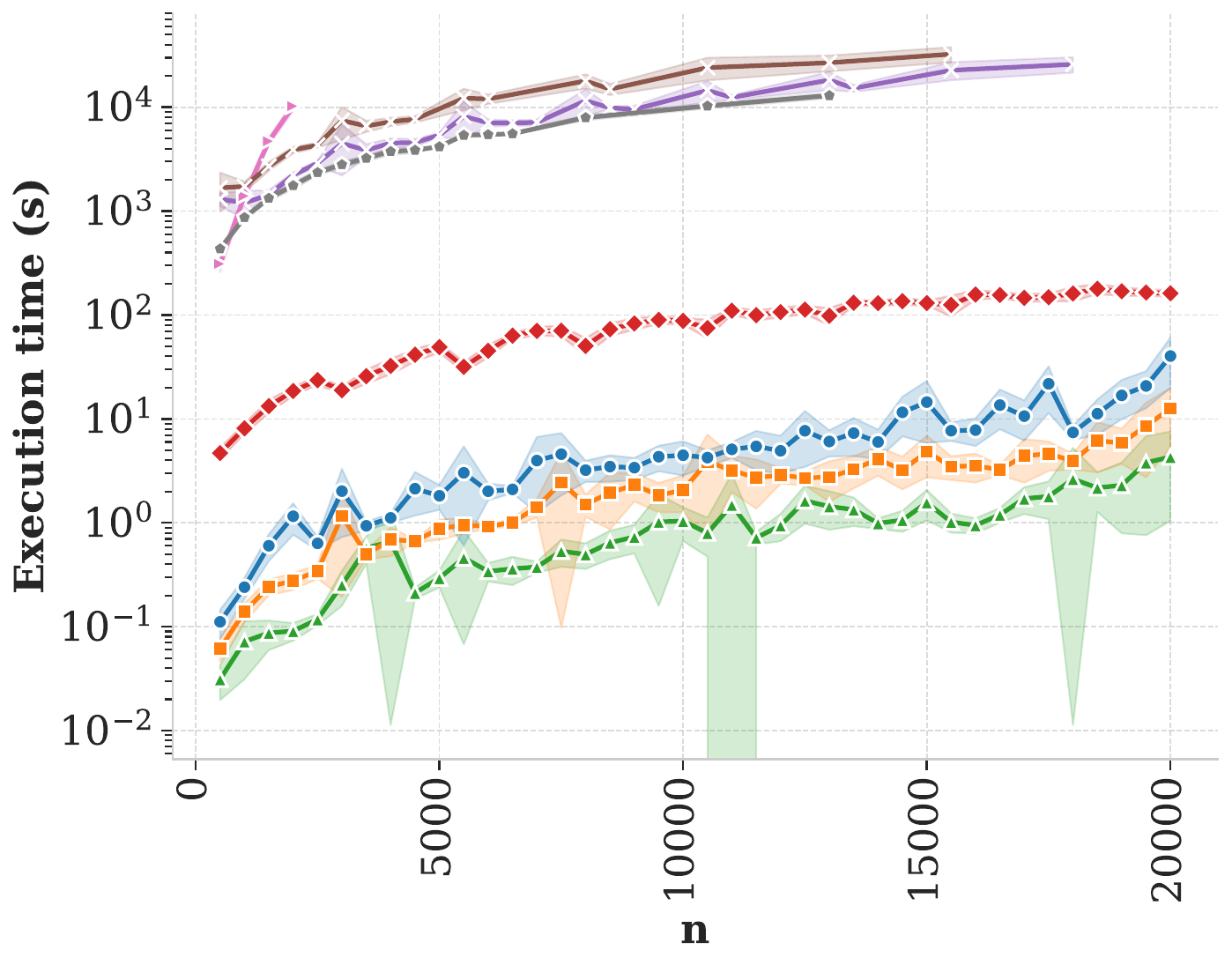} 
    }
    \subfigure{
      \includegraphics[width=0.7\textwidth]{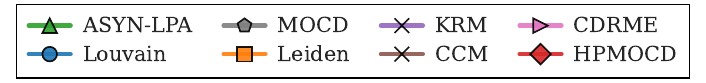} 
    }
    \caption{Scalability analysis on synthetic LFR networks of increasing size ($n$) at fixed mixing parameter $\mu = 0.3$. Metrics include: (a) NMI, (b) AMI, (c) modularity, and (d) runtime (log scale). Results are averaged over 20 runs with 95\% confidence intervals. Gaps indicate methods that exceeded a 32-hour runtime limit.}
    \label{fig:experiment-n-synthetic}
\end{figure}

\autoref{fig:experiment-n-synthetic} presents a comparative evaluation of the scalability of the algorithms as a function of network size. Synthetic LFR benchmark networks were generated with node counts ranging from $[500, \ldots, 20{,}000]$, with fixed $\mu = 0.3$ to maintain moderate structural clarity. Each result corresponds to the average over 20 independent runs with different seeds, and includes 95\% confidence intervals to quantify variability across runs.

To ensure a feasible and focused evaluation, we excluded MOGA-Net from this experiment. The aforementioned report testing revealed (i) uncompetitive detection performance at $\mu = 0.3$; (ii) excessive execution times, frequently exceeding 24 hours on large networks; and (iii) reliance on an older evolutionary architecture that has been superseded by more efficient paradigms. Removing MOGA-Net enabled clearer comparative viewing into modern methods without compromising reproducibility or computational tractability.

As illustrated in \autoref{fig:experiment-n-synthetic}(a) and (b), HP-MOCD consistently achieves high detection accuracy across all network sizes. Its NMI remains near 1.0, and AMI exceeds 0.9, even for the largest networks tested, demonstrating remarkable robustness to increasing problem dimensionality. In contrast, other MOEA-based methods show substantial drops in accuracy at scale. For example, MOCD exhibits clear stagnation in AMI beyond 5{,}000 nodes, highlighting its limited capacity to generalize as the solution space grows. These results validate the scalability of HP-MOCD’s search and variation strategies, particularly its efficient genetic operators and seeding mechanism.

\autoref{fig:experiment-n-synthetic}(c) reports modularity values as a function of network size. HP-MOCD maintains strong modularity performance throughout, achieving values comparable to Louvain and Leiden, even though it does not directly optimize this metric. This suggests that the Pareto-optimal solutions found by HP-MOCD are not only diverse but also structurally coherent. In contrast, MOCD and other methods fall short, especially as network size increases, indicating difficulty in sustaining intra-community density under scale.

Runtime comparisons in \autoref{fig:experiment-n-synthetic}(d) reveal further advantages. The vertical axis is presented in logarithmic scale to capture the wide variation in execution times. While Louvain, Leiden, and ASYN-LPA remain the fastest overall, HP-MOCD achieves a highly favorable trade-off. That is, it runs orders of magnitude faster than any MOEA competitors while delivering much higher accuracy. Its runtime scales smoothly with network size, thanks to the algorithm’s parallel architecture and linear complexity of its core operations (see Section~\ref{sec:proposed_algorithm}).

To further examine the scalability of HP-MOCD, we conducted a controlled multithreaded performance evaluation using synthetic datasets and varying thread counts. These experiments, detailed in~\autoref{sec:appendixB}, demonstrate that parallelism is not just beneficial but essential for achieving practical runtimes at scale. The greedy heuristics (Louvain, Leiden, Label Propagation) were faster, produced higher modularity, but mostly produced lower quality partitions when compared with ~\methodname, which was confirmed by the NMI and AMI values when compared with the ground truth of the network. 

Moreover, the results show that HP-MOCD achieves near-linear speedup up to 8 threads and continues to benefit moderately up to 16 threads. This confirms that the algorithm's core components, such as selection, variation, and evaluation, parallelize effectively, yielding practical execution times on modern multi-core systems. In comparison, while Louvain and Leiden remain faster in absolute terms due to their greedy design, they consistently produce lower-quality partitions in terms of AMI and NMI, reinforcing the superior trade-off achieved by HP-MOCD between structural quality and computational performance.

In terms of computational scalability, HP-MOCD remains efficient in all tests. For 20{,}000 nodes completing all runs under ($148.36 \pm 12.94$) seconds while maintaining NMI $> 0.9$. Statistically, it outperforms all MOEA baselines in both quality and runtime for larger networks. Louvain and Leiden are faster but deliver inferior clustering quality in AMI and F1-score, placing HP-MOCD as a superior method when both accuracy and tractability are required. Overall, these results confirm that HP-MOCD is well-suited for large-scale community detection tasks. It achieves a consistent balance between detection accuracy, structural quality, and computational efficiency, making it a scalable and practical alternative to both traditional heuristics and evolutionary approaches. 

\subsection{Performance on Real-World Networks}

\subsubsection{Networks Overview}

\autoref{tab:dataset_stats} presents the fourteen real-world networks used in our experimental evaluation. These datasets span a wide range of domains, including social media, e-commerce, scientific literature, technological infrastructures, and biological systems, ensuring a comprehensive assessment of algorithmic performance across structurally diverse networks.
\begin{table}[!ht]
\centering
\caption{Statistics and descriptions of the real-world networks used in this study. Each dataset includes node and edge counts, number of ground-truth communities, and its reference. Domains span social, technological, biological, and academic systems.}
\label{tab:dataset_stats}
\begin{tabular}{p{9cm} c c c c}
\toprule
\textbf{Name and Description} & \textbf{$V$} & \textbf{$E$} & \textbf{$k$} & \textbf{Reference} \\ \midrule
\textbf{Youtube} – User interaction network on a video-sharing platform. & 
1{,}134{,}890 & 2{,}987{,}624  & 8{,}385  & \cite{yang2012defining} \\ \midrule
\textbf{Amazon} – Product co-purchase network from e-commerce transactions. & 
334{,}863  &   925{,}872     & 75{,}149 & \cite{yang2012defining} \\ \midrule
\textbf{AS Internet topology (AS)} – Autonomous systems network snapshot (CAIDA, June 2009). & 
23,752 & 58,416 & 176 & \cite{boguna2010sustaining} \\ \midrule
\textbf{Cora Citation Network (Cora)} – Scientific citation graph among academic publications. & 
23,166 & 89,157 & 70 & \cite{vsubelj2013model} \\ \midrule
\textbf{CiteSeer} – Citation network from the CiteSeer digital library. & 
3,327 & 4,676 & 6 & \cite{bollacker1998citeseer} \\ \midrule
\textbf{Programming (Prog.)} – Citation network of programming-related papers. & 
3,109 & 10,564 & 9 & \cite{ivashkin2016logarithmic} \\ \midrule
\textbf{OS} – Scientific publications in Operating Systems, linked by citations. & 
2,176 & 8,731 & 4 & \cite{ivashkin2016logarithmic} \\ \midrule
\textbf{Network} – Papers related to computer networking and their citation links. & 
1,249 & 4,022 & 4 & \cite{ivashkin2016logarithmic} \\ \midrule
\textbf{Databases} – Scientific articles on database research and citations. & 
1,046 & 3,186 & 7 & \cite{ivashkin2016logarithmic} \\ \midrule
\textbf{Mail Eu-core (Eu-core))} – Email communications from a European research institution. & 
1,005 & 16,706 & 42 & \cite{leskovec2007graph} \\ \midrule
\textbf{American College Football (Football)} – Match network among U.S. college football teams (2000 season). & 
115 & 613 & 12 & \cite{newman2004finding} \\ \midrule
\textbf{Books about US Politics (Polbooks)} – Co-purchase network of political books on Amazon. & 
105 & 441 & 3 & \cite{newman2006modularity} \\ \midrule
\textbf{Dolphins} – Social interactions among dolphins in New Zealand. & 
62 & 159 & 2 & \cite{lusseau2003bottlenose} \\ \midrule
\textbf{Zachary’s Karate Club (Karate)} – Friendship network in a university karate club. & 
34 & 78 & 2 & \cite{zachary1977information} \\ 
\bottomrule
\end{tabular}
\end{table}

The node ($V$) and edge ($E$) counts, number of ground-truth communities ($k$), and bibliographic references are provided for each network. These networks include small, well-known benchmark networks such as \texttt{karate} and \texttt{dolphins}, as well as large-scale graphs such as \texttt{youtube}, \texttt{amazon}, and \texttt{cora}, featuring hundreds of thousands to over a million nodes. Most datasets were sourced from the repository compiled by \cite{ivashkin2016logarithmic} and the SNAP collection from \cite{yang2012defining}, which offer ground-truth community labels for reliable external validation. This diversity in structure, size, and domain is essential for evaluating the scalability and generalization capacity of community detection methods in real-world applications.

\subsubsection{Comparative Performance}
\label{sec:real_comparative}
After validating HP-MOCD’s effectiveness and scalability on controlled synthetic benchmarks, we now assess its behavior in real-world scenarios with complex and noisy structures. To enable a thorough and interpretable evaluation, we organize the comparative analysis of \methodname~across three fronts: (i) performance against other MOEA-based algorithms on small and medium-sized datasets; (ii) performance on large-scale real-world graphs, where most competitors fail to scale; and (iii) a benchmark against traditional single-objective community detection heuristics. All results represent averages over 20 independent runs and include standard deviations. Again, statistical significance is assessed using paired t-tests with $\alpha = 0.05$.

\autoref{tab:moea_comparison-real} reports the average AMI, NMI, modularity, and F1-score for each algorithm on small and medium-sized networks, including \texttt{Karate}, \texttt{Polbooks}, \texttt{Dolphins}, \texttt{Eu-core}, and \texttt{Football}. \methodname~emerges as a consistent top performer in these networks, often achieving the best or statistically comparable results across multiple metrics.
\begin{table}[!ht]
\centering
\caption{Comparison of multi-objective community detection algorithms on real-world datasets. Reported values are mean $\pm$ standard deviation across multiple runs. Boldface indicates statistically significant best performance ($\alpha = 0.05$). Missing entries denote methods that failed to complete within 15 hours.}
\label{tab:moea_comparison-real}
\resizebox{.82\textwidth}{!}{
\begin{tabular}{c|c|c|c|c|c}
\toprule
\textbf{Dataset} & \textbf{Method} & \textbf{AMI} & \textbf{NMI} & \textbf{Modularity} & \textbf{F1-Score} \\
\midrule
\multirow{4}{*}{Databases} & CCM & 0.260 $\pm$ 0.000 & 0.392 $\pm$ 0.000 & 0.510 $\pm$ 0.000 & 0.123 $\pm$ 0.000 \\
 & CDRME & 0.244 $\pm$ 0.019 & 0.269 $\pm$ 0.023 & 0.667 $\pm$ 0.030 & \textbf{0.246 $\pm$ 0.019} \\
 & KRM & 0.240 $\pm$ 0.000 & \textbf{0.395 $\pm$ 0.000} & 0.515 $\pm$ 0.000 & 0.130 $\pm$ 0.000 \\
 & HPMOCD & \textbf{0.313 $\pm$ 0.007} & 0.369 $\pm$ 0.006 & 0.683 $\pm$ 0.023 & 0.171 $\pm$ 0.032 \\
\midrule
\multirow{4}{*}{Network} & CCM & 0.171 $\pm$ 0.000 & 0.268 $\pm$ 0.000 & 0.450 $\pm$ 0.000 & 0.105 $\pm$ 0.000 \\
 & CDRME & 0.228 $\pm$ 0.023 & 0.245 $\pm$ 0.023 & 0.612 $\pm$ 0.023 & \textbf{0.246 $\pm$ 0.042} \\
 & KRM & 0.166 $\pm$ 0.000 & 0.276 $\pm$ 0.000 & 0.435 $\pm$ 0.000 & 0.093 $\pm$ 0.000 \\
 & HPMOCD & \textbf{0.254 $\pm$ 0.014} & \textbf{0.285 $\pm$ 0.012} & \textbf{0.633 $\pm$ 0.020} & 0.176 $\pm$ 0.040 \\
\midrule
\multirow{4}{*}{OS} & CCM & 0.205 $\pm$ 0.000 & 0.296 $\pm$ 0.000 & 0.373 $\pm$ 0.000 & 0.213 $\pm$ 0.000 \\
 & CDRME & 0.292 $\pm$ 0.021 & 0.303 $\pm$ 0.022 & 0.569 $\pm$ 0.013 & \textbf{0.383 $\pm$ 0.050} \\
 & KRM & 0.190 $\pm$ 0.000 & 0.293 $\pm$ 0.000 & 0.359 $\pm$ 0.000 & 0.215 $\pm$ 0.000 \\
 & HPMOCD & 0.282 $\pm$ 0.011 & 0.308 $\pm$ 0.010 & 0.580 $\pm$ 0.014 & 0.195 $\pm$ 0.063 \\
\midrule
\multirow{4}{*}{Dolphins} & CCM & 0.527 $\pm$ 0.049 & 0.549 $\pm$ 0.045 & 0.518 $\pm$ 0.004 & 0.554 $\pm$ 0.038 \\
 & CDRME & 0.438 $\pm$ 0.250 & 0.447 $\pm$ 0.253 & 0.365 $\pm$ 0.200 & \textbf{0.674 $\pm$ 0.130} \\
 & KRM & 0.489 $\pm$ 0.033 & 0.517 $\pm$ 0.020 & 0.508 $\pm$ 0.008 & 0.560 $\pm$ 0.049 \\
 & HPMOCD & 0.532 $\pm$ 0.074 & 0.550 $\pm$ 0.068 & 0.518 $\pm$ 0.006 & 0.544 $\pm$ 0.077 \\
\midrule
\multirow{4}{*}{Eu-core} & CCM & 0.193 $\pm$ 0.000 & 0.375 $\pm$ 0.000 & 0.188 $\pm$ 0.000 & 0.113 $\pm$ 0.000 \\
 & CDRME & 0.471 $\pm$ 0.057 & 0.507 $\pm$ 0.059 & 0.396 $\pm$ 0.030 & 0.269 $\pm$ 0.055 \\
 & KRM & 0.210 $\pm$ 0.000 & 0.413 $\pm$ 0.000 & 0.207 $\pm$ 0.000 & 0.150 $\pm$ 0.000 \\
 & HPMOCD & 0.494 $\pm$ 0.026 & \textbf{0.563 $\pm$ 0.025} & 0.399 $\pm$ 0.009 & \textbf{0.313 $\pm$ 0.047} \\
\midrule
\multirow{4}{*}{Football} & CCM & 0.807 $\pm$ 0.032 & 0.845 $\pm$ 0.028 & 0.578 $\pm$ 0.007 & 0.721 $\pm$ 0.062 \\
 & CDRME & 0.703 $\pm$ 0.161 & 0.746 $\pm$ 0.154 & 0.522 $\pm$ 0.119 & 0.583 $\pm$ 0.176 \\
 & KRM & 0.825 $\pm$ 0.024 & 0.867 $\pm$ 0.019 & 0.563 $\pm$ 0.016 & 0.774 $\pm$ 0.039 \\
 & HPMOCD & \textbf{0.881 $\pm$ 0.013} & \textbf{0.912 $\pm$ 0.010} & 0.584 $\pm$ 0.009 & \textbf{0.864 $\pm$ 0.023} \\
\midrule
\multirow{4}{*}{Karate} & CCM & 0.679 $\pm$ 0.010 & 0.695 $\pm$ 0.010 & 0.418 $\pm$ 0.002 & 0.716 $\pm$ 0.025 \\
 & CDRME & 0.584 $\pm$ 0.390 & 0.590 $\pm$ 0.388 & 0.257 $\pm$ 0.156 & 0.834 $\pm$ 0.149 \\
 & KRM & 0.669 $\pm$ 0.047 & 0.686 $\pm$ 0.044 & 0.415 $\pm$ 0.000 & 0.734 $\pm$ 0.024 \\
 & HPMOCD & 0.658 $\pm$ 0.046 & 0.676 $\pm$ 0.041 & 0.417 $\pm$ 0.004 & 0.698 $\pm$ 0.033 \\
\midrule
\multirow{4}{*}{Polbooks} & CCM & 0.507 $\pm$ 0.005 & 0.527 $\pm$ 0.001 & 0.524 $\pm$ 0.002 & 0.760 $\pm$ 0.012 \\
 & CDRME & 0.447 $\pm$ 0.125 & 0.466 $\pm$ 0.119 & 0.452 $\pm$ 0.117 & 0.707 $\pm$ 0.090 \\
 & KRM & 0.467 $\pm$ 0.017 & 0.505 $\pm$ 0.015 & 0.491 $\pm$ 0.001 & 0.664 $\pm$ 0.001 \\
 & HPMOCD & 0.502 $\pm$ 0.018 & 0.521 $\pm$ 0.016 & 0.525 $\pm$ 0.002 & 0.743 $\pm$ 0.024 \\
\bottomrule
\end{tabular}}
\end{table}

For instance, in the \texttt{Football} dataset, \methodname~achieves the highest AMI (0.881), NMI (0.912), and F1-score (0.864), with a modularity of 0.584 that is statistically indistinguishable from other methods. Similarly, in the \texttt{Eu-core} and \texttt{Network}, \methodname~ranks first or ties with the best in AMI, NMI, and modularity, demonstrating its capacity to simultaneously preserve structure and align with ground-truth communities. In datasets such as \texttt{Karate} and \texttt{Dolphins}, the performance is competitive and balanced. Although other methods sometimes achieve higher F1-scores or NMI individually, \methodname~yields stable, multi-metric quality without overfitting to a single criterion. These observed trends are corroborated by a statistical analysis of \autoref{tab:moea_comparison-real}, which shows that HP-MOCD achieves statistically superior results in 13 of the 32 metric-dataset combinations. In 14 cases, it performs equivalently to the best baseline, particularly in small or sparse networks such as \texttt{Dolphins} and \texttt{Karate}. Statistically inferior results are observed in 5 cases, notably in the F1-score for \texttt{Polbooks} and \texttt{OS}, where CDRME slightly outperforms it. Overall, HP-MOCD demonstrates dominance or equivalence in approximately 85\% of the evaluations, underscoring its generalization ability across diverse real-world networks.

\autoref{tab:hpmocd_moea_comparison-real} presents results for large-scale networks, in which \methodname~was the \emph{only} MOEA-based method able to complete execution within the 15-hour time budget. These datasets represent some of the most structurally complex and voluminous graphs in the benchmark suite, with millions of nodes and edges in certain cases.
\begin{table}[!ht]
\centering
\caption{Large-scale datasets where HPMOCD was the only MOEA to complete within the 15-hour time limit. Mean $\pm$ standard deviation shown for each metric.}
\label{tab:hpmocd_moea_comparison-real}
\begin{tabular}{c|c|c|c|c|c}
\toprule
\textbf{Dataset} & \textbf{Method} & \textbf{AMI} & \textbf{NMI} & \textbf{Modularity} & \textbf{F1-Score} \\
\midrule
\multirow{1}{*}{Prog.} & HPMOCD & \textbf{0.342 $\pm$ 0.008} & \textbf{0.391 $\pm$ 0.007} & \textbf{0.669 $\pm$ 0.019} & \textbf{0.150 $\pm$ 0.039} \\
\midrule
\multirow{1}{*}{AS} & HPMOCD & \textbf{0.345 $\pm$ 0.010} & \textbf{0.419 $\pm$ 0.008} & \textbf{0.503 $\pm$ 0.016} & \textbf{0.070 $\pm$ 0.018} \\
\midrule
\multirow{1}{*}{CiteSeer} & HPMOCD & \textbf{0.199 $\pm$ 0.004} & \textbf{0.318 $\pm$ 0.003} & \textbf{0.792 $\pm$ 0.013} & \textbf{0.033 $\pm$ 0.009} \\
\midrule
\multirow{1}{*}{Amazon} & HPMOCD & \textbf{0.402 $\pm$ 0.006} & \textbf{0.667 $\pm$ 0.001} & \textbf{0.762 $\pm$ 0.012} & \textbf{0.007 $\pm$ 0.001} \\
\midrule
\multirow{1}{*}{Youtube} & HPMOCD & \textbf{0.301 $\pm$ 0.016} & \textbf{0.538 $\pm$ 0.037} & \textbf{0.660 $\pm$ 0.009} & \textbf{0.033 $\pm$ 0.004} \\
\midrule
\multirow{1}{*}{Cora} & HPMOCD & \textbf{0.441 $\pm$ 0.007} & \textbf{0.524 $\pm$ 0.004} & \textbf{0.660 $\pm$ 0.023} & \textbf{0.145 $\pm$ 0.021} \\
\bottomrule
\end{tabular}
\end{table}

In all instances, \methodname~produced valid outputs and delivered competitive community detection performance across all evaluation metrics. For example, in the \texttt{Youtube} dataset (over one million nodes), \methodname~ achieved AMI = 0.301 and NMI = 0.538, moderate to strong indicators of alignment with ground-truth communities. Similarly, in \texttt{Amazon} and \texttt{CiteSeer}, the modularity scores exceeded 0.75, highlighting the method's ability to capture intrinsic structural features at scale. These results emphasize \methodname's scalability advantages, primarily attributed to its linear-time variation operators and hybrid initialization strategies, which effectively mitigate the computational bottlenecks faced by traditional MOEAs. Unlike prior approaches that fail to converge or time out on large graphs, \methodname~demonstrates consistent robustness without sacrificing detection quality, making it a viable solution for real-world applications involving massive networks.

\subsubsection{Comparison with Single-Objective Baselines}
\begin{table}[!ht]
\centering
\caption{Comparison between HPMOCD and single-objective algorithms on small and medium-sized real-world networks. Reported values are mean $\pm$ standard deviation across multiple runs. Boldface indicates statistically significant best performance ($\alpha = 0.05$). Missing entries denote methods that failed to complete within the 15-hour time limit.}
\label{tab:HPMOCD_vs_single-real_small}
\begin{tabular}{c|c|c|c|c|c}
\toprule
\textbf{Dataset} & \textbf{Method} & \textbf{AMI} & \textbf{NMI} & \textbf{Modularity} & \textbf{F1-Score} \\
\midrule
\multirow{4}{*}{Databases} & Asyn-LPA & 0.310 $\pm$ 0.007 & \textbf{0.390 $\pm$ 0.005} & 0.662 $\pm$ 0.015 & 0.162 $\pm$ 0.023 \\
 & Leiden & 0.321 $\pm$ 0.003 & 0.346 $\pm$ 0.003 & \textbf{0.756 $\pm$ 0.001} & 0.259 $\pm$ 0.005 \\
 & Louvain & 0.319 $\pm$ 0.006 & 0.344 $\pm$ 0.006 & 0.752 $\pm$ 0.002 & 0.258 $\pm$ 0.006 \\
 & HPMOCD & 0.313 $\pm$ 0.007 & 0.369 $\pm$ 0.006 & 0.683 $\pm$ 0.023 & 0.171 $\pm$ 0.032 \\
\midrule
\multirow{4}{*}{Network} & Asyn-LPA & 0.242 $\pm$ 0.005 & 0.292 $\pm$ 0.005 & 0.612 $\pm$ 0.012 & 0.171 $\pm$ 0.034 \\
 & Leiden & \textbf{0.289 $\pm$ 0.012} & \textbf{0.305 $\pm$ 0.011} & \textbf{0.692 $\pm$ 0.002} & 0.259 $\pm$ 0.005 \\
 & Louvain & 0.276 $\pm$ 0.019 & 0.292 $\pm$ 0.019 & 0.681 $\pm$ 0.004 & 0.255 $\pm$ 0.013 \\
  & HPMOCD & 0.254 $\pm$ 0.014 & 0.285 $\pm$ 0.012 & 0.633 $\pm$ 0.020 & 0.176 $\pm$ 0.040 \\
\midrule
\multirow{4}{*}{OS} & Asyn-LPA & 0.295 $\pm$ 0.009 & 0.337 $\pm$ 0.008 & 0.496 $\pm$ 0.049 & \textbf{0.428 $\pm$ 0.090} \\
 & Leiden & 0.327 $\pm$ 0.008 & 0.340 $\pm$ 0.008 & \textbf{0.641 $\pm$ 0.004} & 0.295 $\pm$ 0.026 \\
 & Louvain & 0.322 $\pm$ 0.007 & 0.335 $\pm$ 0.007 & 0.631 $\pm$ 0.003 & 0.311 $\pm$ 0.021 \\
 & HPMOCD & 0.282 $\pm$ 0.011 & 0.308 $\pm$ 0.010 & 0.580 $\pm$ 0.014 & 0.195 $\pm$ 0.063 \\
\midrule
\multirow{4}{*}{Dolphins} & Asyn-LPA & 0.585 $\pm$ 0.201 & 0.604 $\pm$ 0.191 & 0.465 $\pm$ 0.054 & 0.654 $\pm$ 0.194 \\
 & Leiden & 0.506 $\pm$ 0.053 & 0.523 $\pm$ 0.051 & \textbf{0.524 $\pm$ 0.003} & 0.537 $\pm$ 0.044 \\
 & Louvain & 0.491 $\pm$ 0.044 & 0.508 $\pm$ 0.042 & 0.520 $\pm$ 0.004 & 0.540 $\pm$ 0.048 \\
 & HPMOCD & 0.532 $\pm$ 0.074 & 0.550 $\pm$ 0.068 & 0.518 $\pm$ 0.006 & 0.544 $\pm$ 0.077 \\
\midrule
\multirow{4}{*}{Eu-core} & Asyn-LPA & 0.014 $\pm$ 0.038 & 0.081 $\pm$ 0.038 & 0.013 $\pm$ 0.027 & 0.090 $\pm$ 0.003 \\
 & Leiden & 0.553 $\pm$ 0.010 & 0.589 $\pm$ 0.009 & \textbf{0.434 $\pm$ 0.000} & 0.373 $\pm$ 0.014 \\
 & Louvain & 0.554 $\pm$ 0.011 & 0.590 $\pm$ 0.011 & 0.431 $\pm$ 0.002 & 0.381 $\pm$ 0.027 \\
 & HPMOCD & 0.494 $\pm$ 0.026 & 0.563 $\pm$ 0.025 & 0.399 $\pm$ 0.009 & 0.313 $\pm$ 0.047 \\
\midrule
\multirow{4}{*}{Football} & Asyn-LPA & 0.861 $\pm$ 0.030 & 0.894 $\pm$ 0.025 & 0.587 $\pm$ 0.015 & 0.817 $\pm$ 0.063 \\
 & Leiden & 0.852 $\pm$ 0.018 & 0.883 $\pm$ 0.016 & 0.604 $\pm$ 0.000 & 0.808 $\pm$ 0.039 \\
 & Louvain & 0.834 $\pm$ 0.024 & 0.868 $\pm$ 0.021 & 0.603 $\pm$ 0.005 & 0.774 $\pm$ 0.052 \\
 & HPMOCD & 0.881 $\pm$ 0.013 & 0.912 $\pm$ 0.010 & 0.584 $\pm$ 0.009 & \textbf{0.864 $\pm$ 0.023} \\
\midrule
\multirow{4}{*}{Karate} & Asyn-LPA & 0.478 $\pm$ 0.000 & 0.506 $\pm$ 0.000 & 0.437 $\pm$ 0.000 & 0.661 $\pm$ 0.000 \\
 & Leiden & 0.567 $\pm$ 0.000 & 0.588 $\pm$ 0.000 & 0.445 $\pm$ 0.000 & 0.646 $\pm$ 0.000 \\
 & Louvain & 0.578 $\pm$ 0.000 & 0.600 $\pm$ 0.000 & 0.444 $\pm$ 0.000 & \textbf{0.687 $\pm$ 0.000} \\
 & HPMOCD & 0.558 $\pm$ 0.031 & 0.580 $\pm$ 0.029 & 0.443 $\pm$ 0.004 & 0.647 $\pm$ 0.015 \\
\midrule
\multirow{4}{*}{Polbooks} & Asyn-LPA & 0.501 $\pm$ 0.034 & 0.522 $\pm$ 0.031 & 0.508 $\pm$ 0.021 & 0.747 $\pm$ 0.052 \\
 & Leiden & 0.539 $\pm$ 0.021 & 0.555 $\pm$ 0.019 & \textbf{0.527 $\pm$ 0.000} & 0.760 $\pm$ 0.041 \\
 & Louvain & 0.525 $\pm$ 0.025 & 0.542 $\pm$ 0.024 & 0.527 $\pm$ 0.000 & 0.751 $\pm$ 0.049 \\
 & HPMOCD & 0.502 $\pm$ 0.018 & 0.521 $\pm$ 0.016 & 0.525 $\pm$ 0.002 & 0.743 $\pm$ 0.024 \\
\bottomrule
\end{tabular}
\end{table}

\autoref{tab:HPMOCD_vs_single-real_small} shows that, in small and medium-sized real-world networks, HP-MOCD performs comparably to the best single-objective methods across all evaluation metrics. In datasets such as \texttt{Football}, \texttt{Eu-core}, and \texttt{Karate}, HP-MOCD achieves performance that is either statistically equivalent or marginally below the top performers, with variations well within the standard deviations. For instance, in the \texttt{Football} dataset, HP-MOCD reaches the highest AMI and NMI scores and the best F1 score, closely matching the modularity achieved by Louvain and Leiden.

In more challenging cases like \texttt{OS} and \texttt{Databases}, HP-MOCD shows slightly lower F1-scores or modularity than Louvain or Leiden. However, these differences are offset by HP-MOCD's ability to maintain strong external validation metrics (AMI, NMI) and avoid overfitting to a single objective. This balance underscores HP-MOCD's core advantage: its multi-objective formulation enables simultaneous optimization of both structural quality and label fidelity, which is often not achievable with greedy, single-objective baselines.

Overall, HP-MOCD demonstrates robust generalization and stable behavior, offering a principled trade-off between competing objectives in domains where small gains in one metric can often come at the expense of another. These results highlight that even when not always ranking first, HP-MOCD consistently delivers high-quality, multi-dimensional solutions across diverse network structures.

\autoref{tab:HPMOCD_vs_single-real_large} presents the results for large-scale networks, where HP-MOCD's relative performance is more modest compared to single-objective baselines like Leiden and Louvain. In datasets such as \texttt{Amazon}, \texttt{CiteSeer}, \texttt{Youtube}, and \texttt{Cora}, these methods—known for directly optimizing modularity and their computational efficiency—frequently attain the best modularity and F1-scores, especially in networks exhibiting clear community structures.

\begin{table}[!ht]
\centering
\caption{Comparison between HPMOCD and single-objective algorithms on large-scale real-world datasets. Reported values are mean $\pm$ standard deviation across multiple runs. Boldface indicates statistically significant best performance ($\alpha = 0.05$). Missing entries denote methods that failed to complete within the 15-hour time limit.}
\label{tab:HPMOCD_vs_single-real_large}
\begin{tabular}{@{}p{1cm}|p{1.5cm}|c|c|c|c@{}}
\toprule
\textbf{Dataset} & \textbf{Method} & \textbf{AMI} & \textbf{NMI} & \textbf{Modularity} & \textbf{F1-Score} \\
\midrule
\multirow{4}{*}{Prog.} & Asyn-LPA & 0.342 $\pm$ 0.006 & \textbf{0.417 $\pm$ 0.005} & 0.650 $\pm$ 0.013 & 0.153 $\pm$ 0.027 \\
 & Leiden & \textbf{0.387 $\pm$ 0.006} & 0.402 $\pm$ 0.006 & \textbf{0.759 $\pm$ 0.001} & 0.289 $\pm$ 0.006 \\
 & Louvain & 0.377 $\pm$ 0.006 & 0.392 $\pm$ 0.006 & 0.750 $\pm$ 0.002 & 0.290 $\pm$ 0.005 \\
 & HPMOCD & 0.342 $\pm$ 0.008 & 0.391 $\pm$ 0.007 & 0.669 $\pm$ 0.019 & 0.150 $\pm$ 0.039 \\
\midrule
\multirow{4}{*}{AS} & Asyn-LPA & 0.265 $\pm$ 0.035 & 0.354 $\pm$ 0.045 & 0.371 $\pm$ 0.073 & 0.155 $\pm$ 0.087 \\
 & Leiden & \textbf{0.480 $\pm$ 0.002} & \textbf{0.492 $\pm$ 0.002} & \textbf{0.644 $\pm$ 0.001} & 0.286 $\pm$ 0.009 \\
 & Louvain & 0.475 $\pm$ 0.005 & 0.487 $\pm$ 0.004 & 0.632 $\pm$ 0.002 & \textbf{0.315 $\pm$ 0.030} \\
  & HPMOCD & 0.345 $\pm$ 0.010 & 0.419 $\pm$ 0.008 & 0.503 $\pm$ 0.016 & 0.070 $\pm$ 0.018 \\
\midrule
\multirow{4}{*}{CiteSeer} & Asyn-LPA & 0.184 $\pm$ 0.002 & \textbf{0.347 $\pm$ 0.001} & 0.710 $\pm$ 0.008 & 0.017 $\pm$ 0.002 \\
 & Leiden & \textbf{0.239 $\pm$ 0.002} & 0.330 $\pm$ 0.002 & \textbf{0.895 $\pm$ 0.001} & 0.110 $\pm$ 0.006 \\
 & Louvain & 0.237 $\pm$ 0.003 & 0.328 $\pm$ 0.002 & 0.891 $\pm$ 0.001 & 0.106 $\pm$ 0.006 \\
 & HPMOCD & 0.199 $\pm$ 0.004 & 0.318 $\pm$ 0.003 & 0.792 $\pm$ 0.013 & 0.033 $\pm$ 0.009 \\
\midrule
\multirow{4}{*}{Amazon} & Asyn-LPA & 0.368 $\pm$ 0.000 & \textbf{0.680 $\pm$ 0.000} & 0.708 $\pm$ 0.000 & 0.004 $\pm$ 0.000 \\
 & Leiden & \textbf{0.493 $\pm$ 0.000} & 0.572 $\pm$ 0.000 & \textbf{0.932 $\pm$ 0.000} & \textbf{0.171 $\pm$ 0.000} \\
 & Louvain & 0.479 $\pm$ 0.000 & 0.558 $\pm$ 0.000 & 0.926 $\pm$ 0.000 & 0.164 $\pm$ 0.000 \\
 & HPMOCD & 0.402 $\pm$ 0.006 & 0.667 $\pm$ 0.001 & 0.762 $\pm$ 0.012 & 0.007 $\pm$ 0.001 \\
\midrule
\multirow{4}{*}{Youtube} & Asyn-LPA & 0.311 $\pm$ 0.000 & \textbf{0.582 $\pm$ 0.000} & 0.646 $\pm$ 0.000 & 0.036 $\pm$ 0.000 \\
 & Leiden & 0.307 $\pm$ 0.000 & 0.455 $\pm$ 0.000 & \textbf{0.732 $\pm$ 0.000} & \textbf{0.041 $\pm$ 0.000} \\
 & Louvain & 0.296 $\pm$ 0.000 & 0.444 $\pm$ 0.000 & 0.719 $\pm$ 0.000 & 0.041 $\pm$ 0.000 \\
 & HPMOCD & 0.301 $\pm$ 0.016 & 0.538 $\pm$ 0.037 & 0.660 $\pm$ 0.009 & 0.033 $\pm$ 0.004 \\
\midrule
\multirow{4}{*}{Cora} & Asyn-LPA & 0.417 $\pm$ 0.003 & \textbf{0.552 $\pm$ 0.002} & 0.635 $\pm$ 0.010 & 0.143 $\pm$ 0.020 \\
 & Leiden & \textbf{0.466 $\pm$ 0.005} & 0.474 $\pm$ 0.005 & \textbf{0.800 $\pm$ 0.001} & \textbf{0.238 $\pm$ 0.006} \\
 & Louvain & 0.448 $\pm$ 0.005 & 0.456 $\pm$ 0.005 & 0.790 $\pm$ 0.001 & 0.224 $\pm$ 0.008 \\
 & HPMOCD & 0.441 $\pm$ 0.007 & 0.524 $\pm$ 0.004 & 0.660 $\pm$ 0.023 & 0.145 $\pm$ 0.021 \\
\bottomrule
\end{tabular}
\end{table}

Nevertheless, HP-MOCD remains competitive in several key aspects. It surpasses Asyn-LPA in nearly all metrics, and in datasets like \texttt{Prog.} and \texttt{AS}, its AMI and NMI scores are competitive, showcasing its alignment with ground-truth labels even in complex topologies. Importantly, \methodname~produces these results without relying on greedy heuristics or single-objective optimization, instead leveraging its hybrid initialization and linear-time variation operators to explore a richer solution space. 

Furthermore, the Pareto-based approach provides domain experts with multiple high-quality solutions to choose from, rather than forcing acceptance of a single partition optimized for one criterion. This flexibility is particularly valuable in applications where different stakeholders may prioritize different aspects of community structure, such as cohesion versus separation, or structural versus semantic similarity. The hybrid initialization and linear-time variation operators enable \methodname~to explore solution spaces that greedy single-objective methods might overlook, potentially discovering non-trivial community structures that better reflect the true underlying organization of complex networks.

The consistency observed across all tested datasets underscores \methodname's robustness and reliability. While modularity-based methods suffer from well-known resolution limits that can miss small communities or inappropriately merge distinct ones, \methodname's multi-objective framework naturally addresses these limitations by considering multiple structural perspectives simultaneously. This approach is especially beneficial when ground-truth communities do not strictly align with modularity assumptions, as evidenced by competitive AMI and NMI scores across diverse network topologies. In essence, \methodname~trades peak single-metric performance for a more comprehensive and interpretable understanding of network community structure.

In short, although HP-MOCD does not dominate in large-scale settings against single-objective methods, it offers a more interpretable and generalizable approach by balancing structural and semantic quality. These properties are especially advantageous in real-world applications that demand robustness, fairness, and fidelity across diverse evaluation perspectives.

\subsection{Summary of Findings}

Our evaluation of \methodname~ across synthetic and real-world networks reveals several key insights into its performance characteristics and practical applicability. The algorithm demonstrated exceptional scalability, being the only evolutionary method capable of processing graphs with tens of thousands of nodes within the 15-hour computational budget. In these scalability tests, HP-MOCD outperformed baselines in $34.1\%$ of cases for modularity with an average improvement of $111.91\%$, while achieving superior performance in $74.3\%$ of cases for both NMI and AMI, with average improvements of $50.86\%$ and $113.24\%$ respectively. These results validate the effectiveness of our parallel-ready architecture and optimized genetic operators, which enable practical application to large-scale networks where traditional MOEAs fail.

The robustness experiments further highlighted \methodname~'s resilience to structural noise and ambiguous community boundaries. When tested under varying noise conditions, the algorithm exhibited remarkable stability, outperforming baselines in $69.0\%$ of cases for modularity with an impressive average improvement of $227.93\%$. Similarly strong results were observed for NMI and AMI, where HP-MOCD was superior in $63.8\%$ and $67.2\%$ of cases respectively, achieving average improvements of $144.88\%$ and $230.94\%$. The algorithm's performance degraded gracefully with increasing mixing parameters, maintaining meaningful community detection even when structural boundaries became highly ambiguous. This robustness is particularly valuable for real-world applications where network data often contains noise, missing edges, or uncertain connections.

On real-world networks, HP-MOCD demonstrated competitive performance against established single-objective methods like Louvain and Leiden, achieving balanced results across multiple evaluation criteria. The algorithm outperformed baselines in $55.0\%$ of cases for both modularity and AMI, with average improvements of $58.35\%$ and $65.60\%$ respectively, while being superior in $51.7\%$ of cases for NMI with an average improvement of $13.30\%$. Particularly noteworthy was its performance on small to medium-sized networks such as \texttt{Karate}, \texttt{Dolphins}, and \texttt{Network}, where \methodname~ matched or exceeded the performance of single-objective methods across all metrics with negligible statistical differences. This demonstrates that the multi-objective formulation does not compromise performance on simpler instances and can effectively generalize across different scales.

For large-scale real-world datasets including \texttt{CiteSeer}, \texttt{AS}, and \texttt{Cora}, \methodname~ exhibited interesting trade-off characteristics that highlight its unique advantages. While single-objective methods occasionally achieved higher modularity or F1-scores through greedy optimization of specific objectives, \methodname~ consistently provided more balanced solutions that performed well across multiple evaluation criteria simultaneously. This balance is crucial in practical applications where both structural coherence and alignment with ground-truth communities are valued. Moreover, \methodname's ability to generate a Pareto front of diverse, high-quality solutions in a single run provides users with the flexibility to select solutions based on their specific application requirements, a capability that single-objective methods inherently lack.

The consistent accuracy of \methodname~ across structurally diverse graphs, as evidenced by strong performance in NMI, AMI, and F1-score metrics, indicates that the hybrid multi-objective design successfully balances structural quality and external similarity. This allows the algorithm to generalize effectively even when network characteristics such as degree distribution, density, and number of communities vary significantly. By avoiding overfitting to single metrics, \methodname~ produces well-rounded solutions that better reflect the multifaceted nature of community structure in real networks. These results demonstrate that \methodname~ successfully integrates computational efficiency with robust multi-objective optimization, addressing the limitations of both classical heuristics and traditional MOEAs. The findings underscore the potential of hybrid evolutionary frameworks in tackling complex network analysis problems where multiple, often conflicting objectives must be balanced, making the proposed method a practical and effective solution for modern community detection challenges.


\section{Conclusion and Future Work}\label{sec:conclusion}

We introduced HP-MOCD, a high-performance evolutionary multi-objective algorithm for large-scale network community detection. Built upon the NSGA-II framework, HP-MOCD incorporates topology-aware crossover and mutation operators, coupled with a parallelized and efficient implementation that ensures practical scalability. In contrast to traditional single-objective heuristics and earlier MOEA-based methods, HP-MOCD produces a Pareto front of diverse, high-quality partitions, enabling a principled exploration of trade-offs between structural cohesion and ground-truth alignment.

Extensive experiments on synthetic and real-world datasets show that HP-MOCD consistently outperforms prior evolutionary methods in both runtime and robustness, while delivering competitive accuracy, particularly in scenarios involving noisy, ambiguous, or large-scale networks. Even in cases where single-objective methods such as Louvain or Leiden yield higher scores in individual metrics, HP-MOCD maintains a more balanced performance profile across AMI, NMI, modularity, and F1-score, highlighting the benefits of its multi-objective formulation.

Future work can extend the algorithm to support overlapping communities and more general network types, such as weighted, directed, or temporal graphs. Another promising path involves improving robustness under high mixing regimes (\( \mu \to 1 \)), potentially through hybrid strategies or advanced selection mechanisms. An additional direction includes investigating the resilience of communities under node or edge failures, as explored in prior work on failure-resilient communities~\cite{wang2018constructing, wang2019community}. We also envision extending our experimental evaluation by incorporating additional datasets and more baseline algorithms, such as those proposed by~\cite{Ghoshal:2019, Ghoshal:2021}, to more thoroughly assess the strengths and limitations of HP-MOCD in diverse scenarios. Furthermore, accurate and scalable community detection can serve as a foundation for downstream applications such as diffusion source inference and diffusion containment in complex networks, which increasingly rely on community structure to enhance performance~\cite{liu2023containment, liu2024observer, liu2025source}. Lastly, we plan to explore integration with NSGA-III~\cite{nsga3} and other recent MOEA frameworks to support higher-dimensional objectives, such as conductance and community balance, thereby expanding the scope of multi-objective community detection.

\section*{Acknowledgments}
    The authors would also like to thank the \textit{Coordenação de Aperfeiçoamento de Pessoal de Nível Superior} - Brazil (CAPES) - Finance Code 001, \textit{Fundacão de Amparo à Pesquisa do Estado de Minas Gerais} (FAPEMIG, grants APQ-01518-21), \textit{Conselho Nacional de Desenvolvimento Científico e Tecnológico} (CNPq, grants 308400/2022-4) and Universidade Federal de Ouro Preto (PROPPI/UFOP) for supporting the development of this study.

\bibliographystyle{main}  
\bibliography{main}  

\begin{appendices}

\section{Space Complexity Analysis}
\label{sec:appendixA}

Let $|V|$ represent the number of nodes, $|E|$ the number of edges, $N_p$ the population size, and $d_{\max} = \max_{v \in V} d(v)$ the maximum node degree. The graph $G$ itself is stored once, requiring $O\bigl(|V| + |E|\bigr)$ space.
Each individual in the population is a labeling of all $|V|$ nodes, thus one individual occupies $O\bigl(|V|\bigr)$ space.
At any generation, HP-MOCD holds both the parent population $P_t$ and the offspring population $Q_t$. This amounts to $2N_p$ individuals, consuming
\[
O\bigl(2\,N_p \cdot |V|\bigr) \;=\; O\bigl(N_p\,|V|\bigr) \quad \text{space}.
\]

Regarding algorithmic components:
\begin{itemize}
    \item For \emph{Crossover}, a hash-map is employed to count labels dynamically. This map is reused for each node, resulting in $O(1)$ auxiliary space per node, which does not accumulate across nodes.
    \item For \emph{Mutation}, apart from storing the new individual (requiring $O\bigl(|V|\bigr)$), a frequency map of size at most the node's degree ($d(v)$) may be allocated when mutating a single node. Since this map is reused across nodes, the total auxiliary space for this operation is $O(d_{\max})$.
\end{itemize}

The non-dominated sorting and crowding distance calculations (as in NSGA-II) are performed on $2N_p$ individuals. The space required for these operations includes:
\[
\underbrace{O\bigl(N_p\bigr)}_{\substack{\text{domination counts}\\\text{and front indices}}}
\;+\;
\underbrace{O\bigl(N_p\bigr)}_{\substack{\text{crowding}\\\text{distance array}}}
\;=\; O\bigl(N_p\bigr).
\]

\subsection{Total Space Complexity}
The total space complexity is the sum of these components:
\[
\begin{aligned}
   & O\bigl(|V| + |E|\bigr) && \text{(graph storage)} \\
   +\;& O\bigl(N_p\,|V|\bigr)    && \text{(populations } P_t, Q_t \text{)} \\
   +\;& O\bigl(d_{\max}\bigr)   && \text{(mutation auxiliary space)} \\
   +\;& O\bigl(N_p\bigr)       && \text{(NSGA-II auxiliary space)} \\
   \;=\;& O\bigl(N_p\,|V| + |V| + |E| + d_{\max} + N_p\bigr).
\end{aligned}
\]
In typical scenarios involving sparse graphs (where $|E| = O\bigl(|V|\bigr)$) and for realistic algorithm parameters where $N_p\,|V|$ is the overwhelmingly largest term (i.e., $N_p\,|V| \gg |V|, |E|, d_{\max}, N_p$), the dominant space complexity is
\[
\boxed{O\bigl(N_p\,|V|\bigr).}
\]

\section{Role of Multi-Threaded Architecture in Algorithm Performance}\label{sec:appendixB}

The inherent parallel design of our algorithm necessitates an investigation into its scalability with an increasing number of logical threads and the identification of optimal threading levels. Understanding this relationship is paramount to harnessing the algorithm's full potential.

To quantify the impact of multi-threading, we benchmarked the algorithm on a high-performance system featuring an AMD Ryzen Threadripper 3960X processor (24 physical cores, 48 logical threads, @ 3.70GHz) and 128GB DDR4 RAM. The number of execution threads was systematically varied from 1 to 48. These tests utilized the same synthetic graph datasets with \( n \in [10,000, 30,000, 50,000] \) nodes as detailed in our scalability analysis (\autoref{sec:paramters-setup}). Each configuration was executed five times, a sufficient number given the consistently low standard deviation observed in execution times.

\begin{figure}[!th]
 \centering
 \subfigure[10,000 nodes]{ 
 \includegraphics[width=0.3\textwidth]{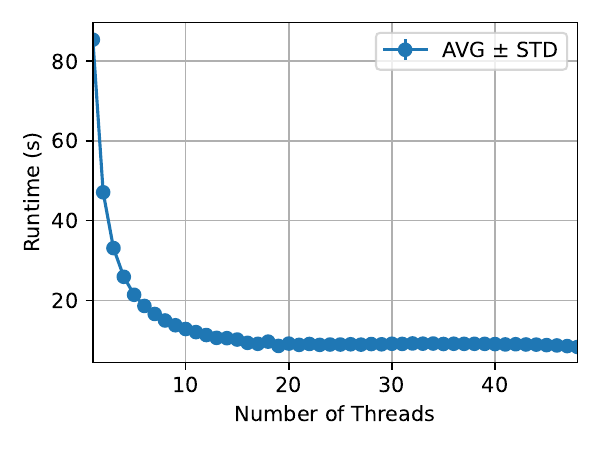}
 }
 \subfigure[30,000 nodes]{ 
 \includegraphics[width=0.3\textwidth]{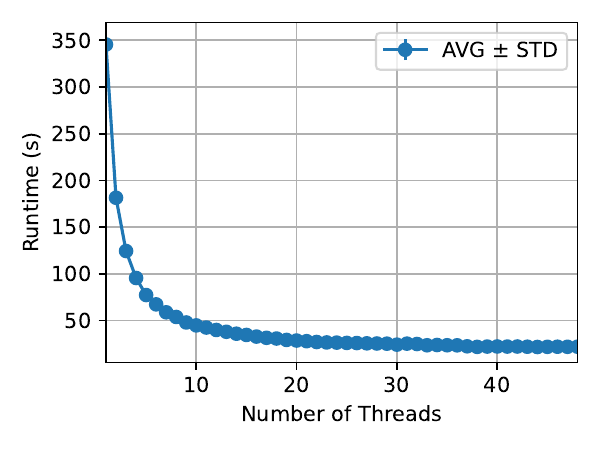}
 }
 \subfigure[50,000 nodes]{ 
 \includegraphics[width=0.3\textwidth]{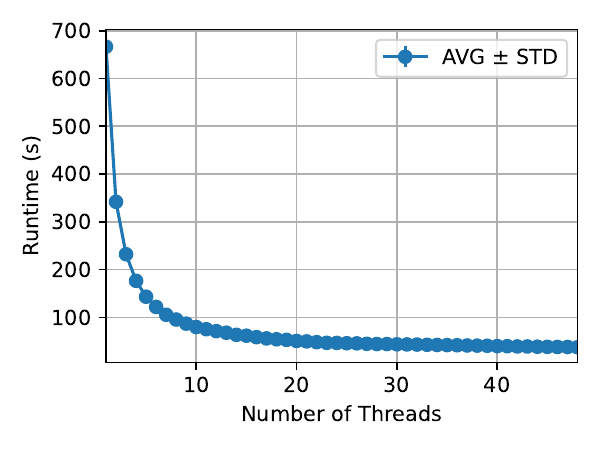}
 }
 \caption{Average runtime ($mean \pm \sigma$) as a function of the number of logical threads for graphs with 10,000 (a), 30,000 (b), and 50,000 (c) vertices. The results underscore the significant speedup achieved through multi-threading compared to single-threaded execution.}
 \label{fig:threads_benchmarks}
\end{figure}

\autoref{fig:threads_benchmarks} illustrates the performance gains. Initially, increasing the number of threads yields substantial reductions in runtime. However, a performance plateau is reached beyond a certain threshold (approximately 8 threads for 10,000 nodes, 15 for 30,000 nodes, and 20 for 50,000 nodes). Beyond these points, adding more threads provides marginal benefits. This is likely attributable to the increased overhead associated with managing a large number of concurrent threads.

The most compelling evidence for the importance of the multi-threaded architecture lies in the stark contrast with single-threaded execution. As shown in \autoref{fig:threads_benchmarks}, a linear (single-thread) approach is dramatically slower across all scenarios. Specifically, when compared to utilizing the maximum available logical threads, single-threaded execution was \textbf{10.27} times slower for 10,000 nodes (\autoref{fig:threads_benchmarks}(a)), \textbf{16.45} for 30,000 nodes (\autoref{fig:threads_benchmarks}(b)), and \textbf{17.83} times for 50,000 nodes (\autoref{fig:threads_benchmarks}(c)). These figures unequivocally demonstrate that the algorithm's parallel processing capability is not merely an optimization but a fundamental requirement for achieving practical execution times, rendering it viable and effective for real-world applications. Without its multi-threaded design, the algorithm would be impractically slow for datasets of meaningful scale.

\section{Selection of Genetic Algorithm Parameters}\label{sec:appendixC}
\label{sec:paramters-setup}
The performance of the NSGA-II algorithm depends critically on two GA parameters: the population size $P$ and the number of generations $G$. These parameters control the total search effort and must be tuned to trade off clustering quality—measured here via AMI and NMI—against computational cost.

\subsection{Aggregate Quality Metric}  
To summarize clustering quality in a single scalar, we compute the harmonic mean of AMI and NMI:
\[
H \;=\; \frac{2 \,\bigl(\mathrm{AMI}\times \mathrm{NMI}\bigr)}{\mathrm{AMI} + \mathrm{NMI}},
\]
which balances the two scores and more strongly penalizes configurations where either AMI or NMI is low. While AMI and NMI are widely used, the methodology remains compatible with other external or internal evaluation metrics, depending on application context.

\subsection{GA Effort Metric}  
As a proxy for total computational investment, we define:
\[
E \;=\; P \times G,
\]
i.e.\ the total number of candidate evaluations across all generations. This single metric captures both population diversity ($P$) and evolutionary depth ($G$).

A systematic grid search was performed over various $(P,G)$ pairs for graph sizes $n \in [10\,000,\;20\,000,\;30\,000;40\,000]$. For each configuration, we recorded $H$ and the execution time.

\begin{figure}[!ht]
 \centering
 \subfigure[H versus E]{ 
 \includegraphics[width=0.45\textwidth]{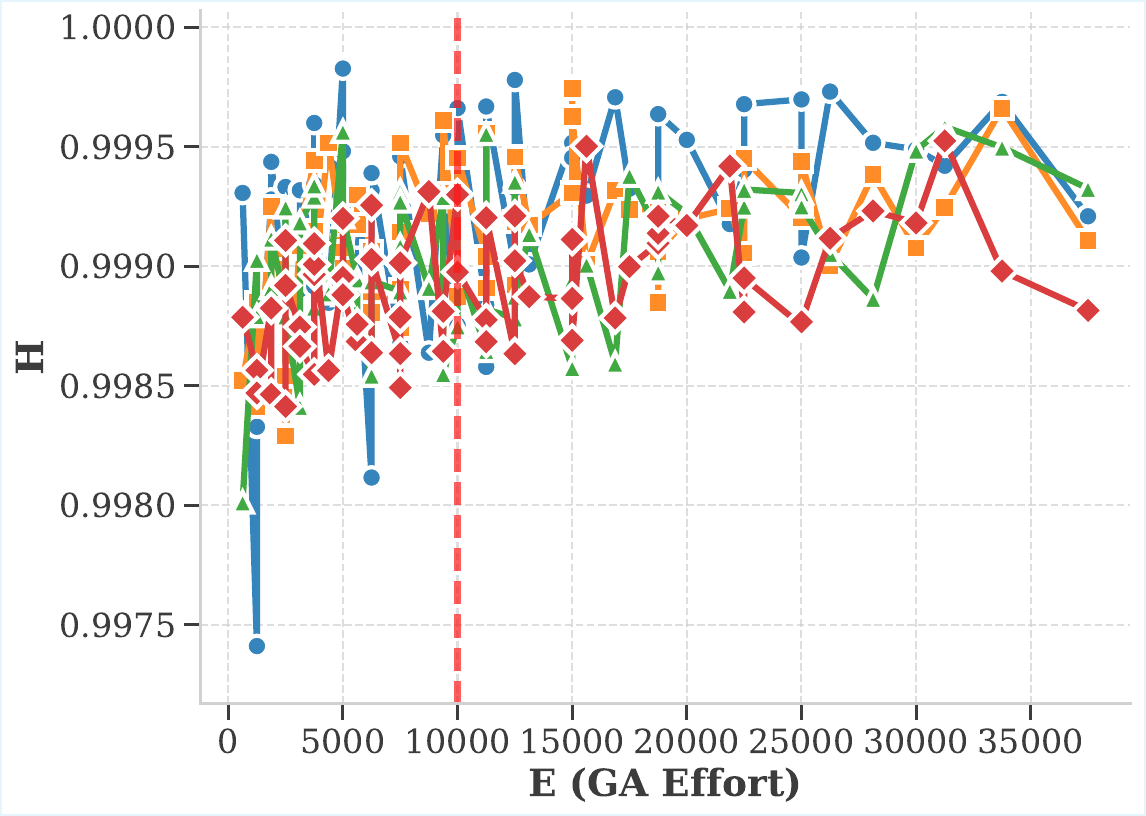}
 }\label{fig:ga_quality_vs_paramA}
 \subfigure[Execution Time versus E]{ 
 \includegraphics[width=0.45\textwidth]{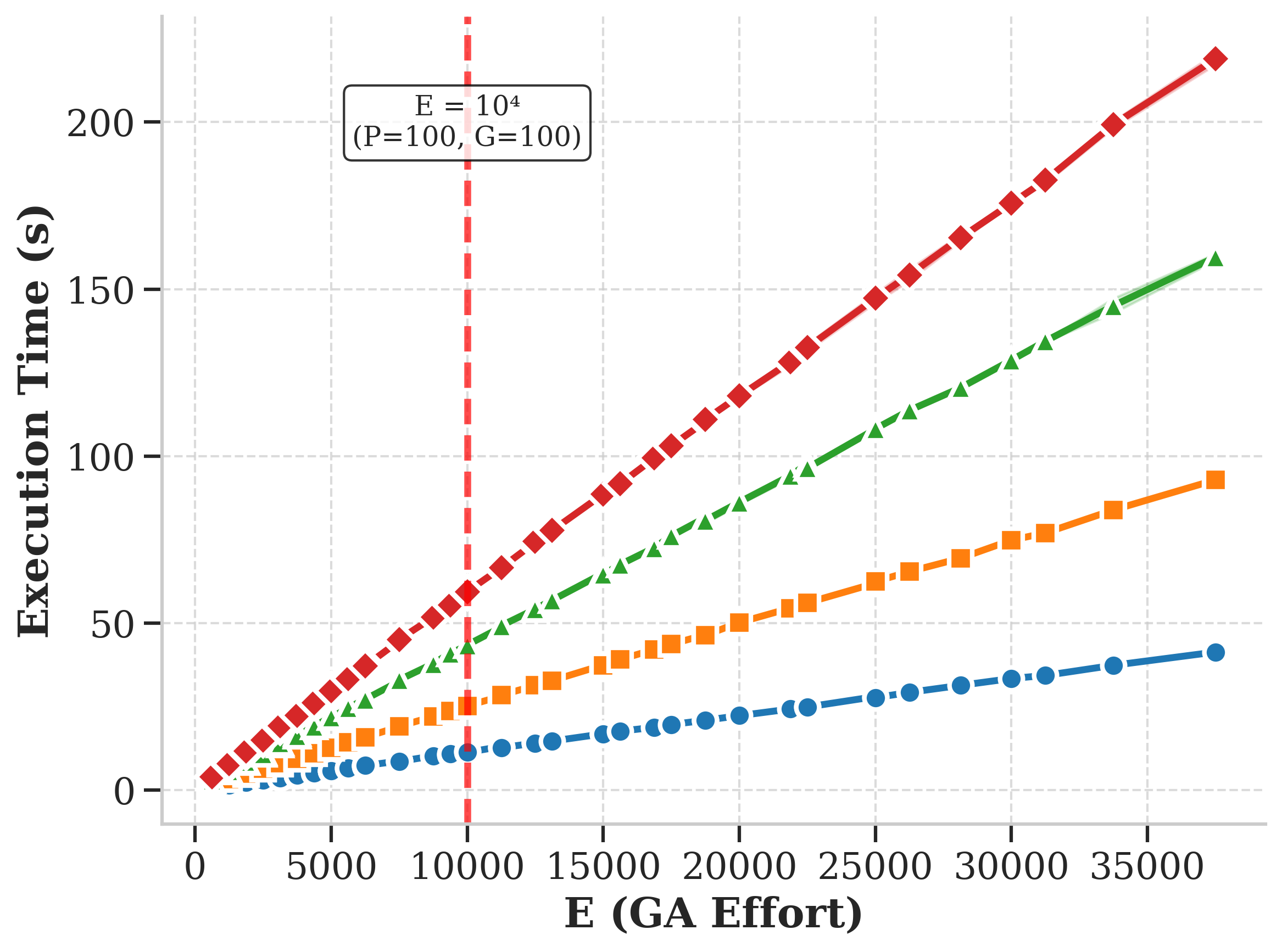}
 }\label{fig:ga_quality_vs_paramB}
  \subfigure{ 
 \includegraphics[width=1\textwidth]{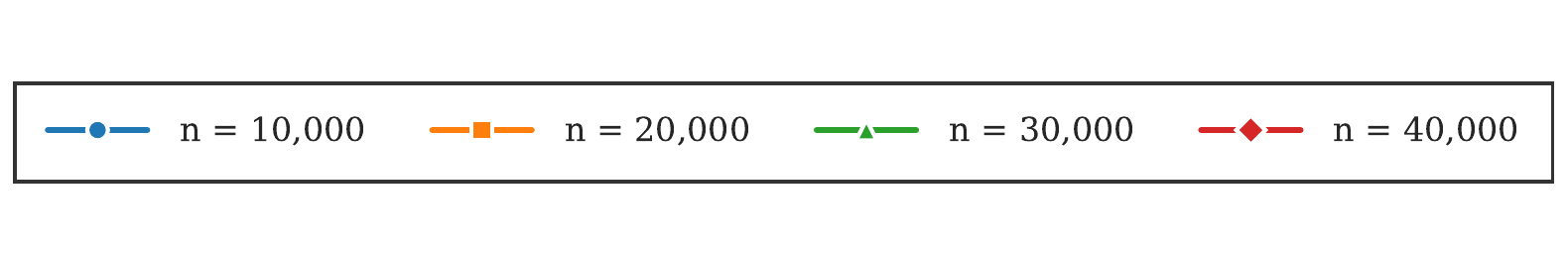}
 }
\caption{Harmonic mean of AMI and NMI versus GA effort $E$, and execution time versus GA effort $E$, for different graph sizes.}
\label{fig:ga_quality_vs_param}
\end{figure}

\autoref{fig:ga_quality_vs_param}(a) and \autoref{fig:ga_quality_vs_param}(b) display, respectively, the harmonic mean quality and runtime as functions of $E$, with separate curves for each $n$.

From these figures, we observe that clustering quality $H$ increases rapidly with $E$ up to approximately $10^4$, after which gains become marginal. Meanwhile, execution time grows roughly linearly with $E$ for all tested graph sizes. This behavior highlights a clear trade-off regime where further computational effort yields diminishing returns.

\subsection{Chosen Configuration}  
To balance near‐peak quality with reasonable runtime, we select:
\[
P = 100,\quad G = 100,\quad (E = 10^4)
\]
as this setting reaches near-optimal $H$ before the saturation point, while keeping execution time moderate and scalable.

\section{Crossover and Mutation Example}\label{sec:appendixE}

\autoref{fig:operator_examples} provides a visual example of the crossover and mutation operators described in Section~\ref{sec:operators}. These examples help illustrate how the genetic variation mechanisms alter individuals during the evolutionary process.

\newgeometry{
    left=1.5cm,
    right=3.5cm,
    top=1.5cm,
    bottom=1.5cm
}

\begin{landscape}
\thispagestyle{mylandscape}
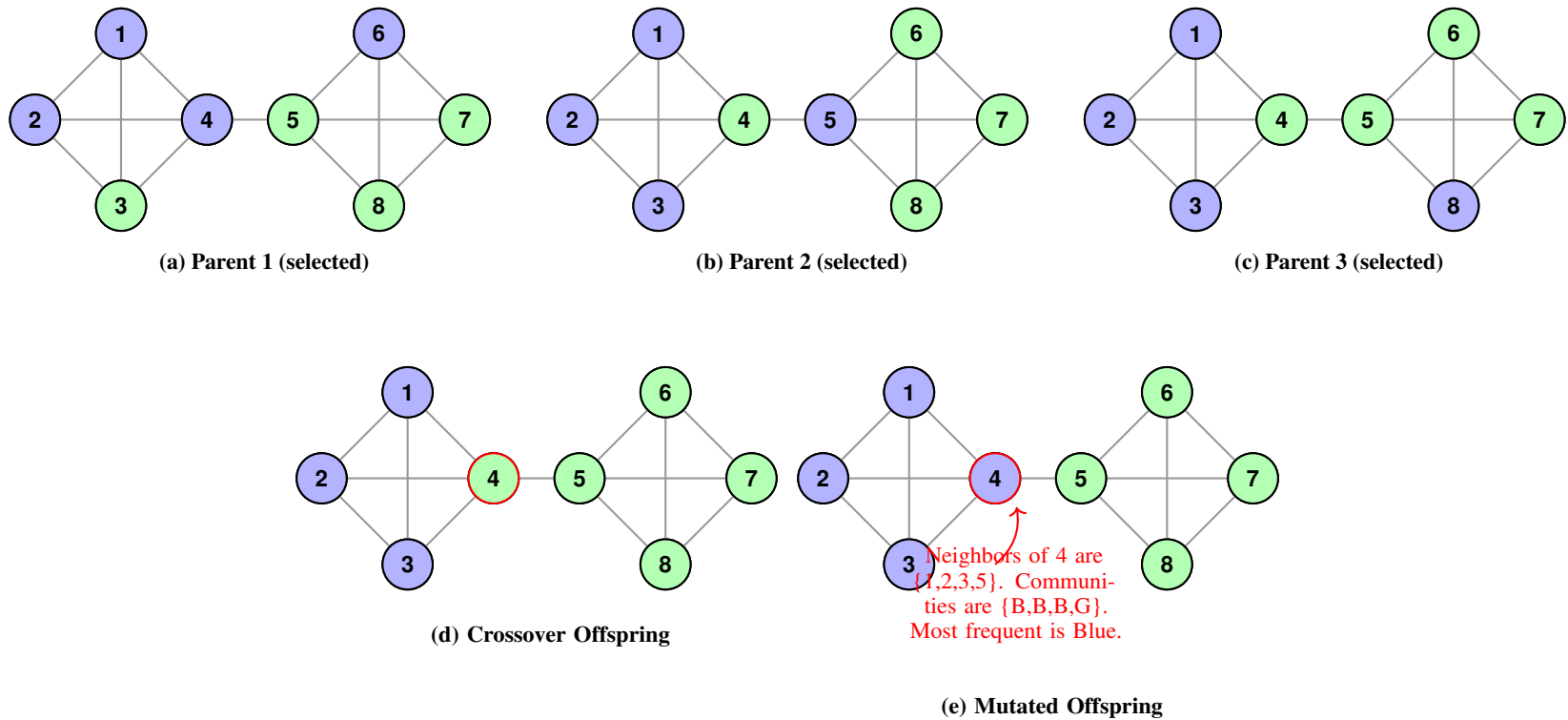
\begin{figure}[h!]
\small
\centering
\begin{tikzpicture}[
    node_style/.style={circle, draw, thick, minimum size=0.7cm, font=\sffamily\bfseries},
    blue_node/.style={node_style, fill=blue!30, text=black},
    green_node/.style={node_style, fill=green!30, text=black},
    edge_style/.style={thick, gray!80},
    highlight_node/.style={node_style, draw=red, line width=1.5pt} 
]

\newcommand{\drawbasegraph}{
    \node[node_style] (n1) at (0,1.2) {1};
    \node[node_style] (n2) at (-1.2,0) {2};
    \node[node_style] (n3) at (0,-1.2) {3};
    \node[node_style] (n4) at (1.2,0) {4};
    \node[node_style] (n5) at (2.4,0) {5};
    \node[node_style] (n6) at (3.6,1.2) {6};
    \node[node_style] (n7) at (4.8,0) {7};
    \node[node_style] (n8) at (3.6,-1.2) {8};

    \graph[edges={edge_style}] {
        (n1) -- (n2) -- (n3) -- (n1);
        (n4) -- {(n1), (n2), (n3)};
        (n6) -- (n7) -- (n8) -- (n6);
        (n5) -- {(n6), (n7), (n8)};
        (n4) -- (n5);
    };
}

\begin{scope}[shift={(-7.5, 5)}]
    \drawbasegraph
    \foreach \i in {1,2,4,6} { \node[blue_node] at (n\i) {\i}; }
    \foreach \i in {3,5,7,8} { \node[green_node] at (n\i) {\i}; }
    \node[font=\bfseries] at (2, -2) {(a) Parent 1 (selected)};
\end{scope}

\begin{scope}[shift={(0, 5)}]
    \drawbasegraph
    \foreach \i in {1,2,3,5} { \node[blue_node] at (n\i) {\i}; }
    \foreach \i in {4,6,7,8} { \node[green_node] at (n\i) {\i}; }
    \node[font=\bfseries] at (2, -2) {(b) Parent 2 (selected)};
\end{scope}

\begin{scope}[shift={(7.5, 5)}]
    \drawbasegraph
    \foreach \i in {1,2,3,8} { \node[blue_node] at (n\i) {\i}; }
    \foreach \i in {4,5,6,7} { \node[green_node] at (n\i) {\i}; }
    \node[font=\bfseries] at (2, -2) {(c) Parent 3 (selected)};
\end{scope}

\begin{scope}[shift={(-3.5, 0)}]
    \drawbasegraph
    \foreach \i in {1,2,3} { \node[blue_node] at (n\i) {\i}; }
    \foreach \i in {4,5,6,7,8} { \node[green_node] at (n\i) {\i}; }
    \node[highlight_node, green_node] at (n4) {4};
    \node[font=\bfseries, text width=5cm, align=center] at (2, -2.2) {(d) Crossover Offspring};
\end{scope}

\begin{scope}[shift={(3.5, 0)}]
    \drawbasegraph
    \foreach \i in {1,2,3,4} { \node[blue_node] at (n\i) {\i}; }
    \foreach \i in {5,6,7,8} { \node[green_node] at (n\i) {\i}; }
    \node[highlight_node, blue_node] at (n4) {4};
    \draw[->, thick, red, bend right] (1.2, -1.2) to node[midway, below, text width=3cm, align=center] {Neighbors of 4 are \{1,2,3,5\}. Communities are \{B,B,B,G\}. Most frequent is Blue.} (1.5, -0.4);
    \node[font=\bfseries, text width=5cm, align=center] at (2, -3.2) {(e) Mutated Offspring};
\end{scope}

\end{tikzpicture}
\caption{An illustrative example of the Crossover and Mutation operators. (a), (b) and (c) show three parent solutions selected. (d) The offspring is created by a majority vote for each node's community assignment, as determined by \autoref{eq:arg_max}. The result is a novel partition, different from all parents. (e) The mutation operator inspects the neighborhood of node 4, finds that blue is the most frequent community, and flips its color to perform a final correction, recovering the optimal partition.}
\label{fig:operator_examples}
\end{figure}
\end{landscape}

\end{appendices}

\end{document}